\documentclass[preprint,12pt,3p]{elsarticle}

\usepackage{booktabs}
\usepackage{makecell}
\usepackage{url}
\usepackage{graphicx}
\usepackage{balance}
\usepackage{amsmath}
\usepackage{subfigure}
\usepackage{graphicx}
\usepackage{balance}
\usepackage{color}
\usepackage{comment}
\usepackage{array}
\usepackage{lipsum}
 \usepackage{caption}
 
\usepackage{blindtext}
\usepackage{enumitem}
\usepackage{lipsum}
\usepackage{capt-of}
\usepackage{tikz}
\usetikzlibrary{arrows,positioning,automata}
\usepackage{balance}
\usepackage{amsmath}
\usepackage[T1]{fontenc}
\usepackage{algpseudocode}
\usepackage{algorithm}
\usepackage{subfigure}
\usepackage{graphicx}
\usepackage[utf8]{inputenc}
\usepackage{balance}
\usepackage{multirow}
\usepackage{color}
\usepackage{comment}
\usepackage{hyperref}
\usepackage{array}
\usepackage{soul}
\usepackage[justification=centering]{caption}
\usepackage[lighttt]{lmodern}
\usepackage{libertine}
\usepackage{amssymb}
\usepackage{float}

\journal{Online Social Networks and Media}

\begin{document}

\begin{frontmatter}


\title{Sentiment Dynamics in Social Media News Channels}

\author[label1]{Nagendra Kumar}
\address[label1]{Indian Institute of Technology Hyderabad, India}


\ead{cs14resch11005@iith.ac.in}


\author[label2]{Rakshita Nagalla}
\address[label2]{Columbia University, New York, USA}
\ead{rn2439@columbia.edu}

\author[label3]{Tanya Marwah}
\address[label3]{Carnegie Mellon University, Pittsburgh, USA}
\ead{tmarwah@andrew.cmu.edu}

\author[label1]{Manish Singh}
\ead{msingh@iith.ac.in}





\begin{abstract}
Social media is currently one of the most important means of news communication. Since people are consuming a large fraction of their daily news through social media, all the traditional news channels are using social media to catch the attention of users. Each news channel has its own strategy to attract more users.  In this paper, we analyze how the news channels use sentiment to garner users' attention in social media. We compare the sentiment of news posts generated by television, radio and print media, to show the differences in the news covered by these channels. We also analyze users' reactions and sentiment of users' opinions on news posts with different sentiments. We do our analysis on the dataset extracted from the Facebook Pages of five popular news channels. Our dataset contains 0.15 million news posts and 1.13 billion users reactions. Our result shows that sentiment of the user opinion strongly correlates with the sentiment of news posts and the type of information source. Our study also illustrates the differences between the social media news channels of different types of news sources. 

\end{abstract}


\end{frontmatter}

\section{Introduction}
\label{sec:sa_int}
Television, radio and print media are the three primary types of news sources in the world. Due to the unique nature of each communication medium and the manner in which their audience consume the information, they show a significant difference in their news articles. Despite these differences, all these sources are commonly accused of using exaggerated headlines to garner attention~\cite{ecker2014effects} and for focusing on negative news~\cite{soroka2015news}. Today, social media has emerged as a powerful platform for consumption of news with 62\% of U.S. adults reportedly getting their news from social media~\cite{gottfried2016news}. Therefore, traditional news channels have started generating and disseminating news through various social media platforms. 

As growing number of people consume, share and discuss news online, it is important to understand whether the lack of regulation inherent in social media is being exploited to spread more aggressive and negative news. We surprisingly found that news is not necessarily negative across all the news channels. Rather, there is considerable variation in the way news is presented by different news channels and is heavily dependent on the medium through which these channels traditionally disseminated news, namely radio, TV or print media. Print and radio media based channels post more positive news while TV based channels post more negative news. The difference is not only because of difference in the type of stories covered but because of how the same news is presented by different types of channels. For instance, we can observe from Table~\ref{table:sa_pon} how the \emph{Dakota Access Pipeline} protest, a major movement in the Northern United States to protect natural resources and spiritual sites, reported by CNN and The Economist are extreme in tone, but opposite in polarity (we measure polarity on the scale of -5 to +5) despite being posted at the same time and referring to the same incident.

\begin{table}[!h]
\begin{center}
\resizebox{\columnwidth}{!}{%
\small
\begin{tabular}{|p{2.5cm}|p{11cm}|p{1.5cm}|}
\hline
\hline
{Organization} & {News} & {Polarity}\\
 \hline\hline
CNN & We're at the Standing Rock Sioux Camp in North Dakota. Protesters here are fighting to block the Dakota Access Pipeline and have vowed to stand their ground — despite growing calls for them to leave camp and threats of prosecution from law enforcement. Any questions for CNN's Sara Sidner? & -5\\
\hline
The Economist & Whatever the final result of the huge, long-running protests by native Americans against the Dakota Access Pipeline, the demonstrations will surely be remembered as a landmark in relations between organised religion, Christianity in particular, and indigenous people & +3\\
\hline
\end{tabular}%
}
 \caption{Polarity of a news generated by two different types of channels}
\label{table:sa_pon}
\end{center}
\end{table}

As opposed to one-to-many communication structure of traditional media, social media enables many-to-many communication by allowing the users to engage with the news articles by liking, sharing and commenting on them. The number of likes, shares, and comments received by a news are good objective measures of engagement and provide insight into what type of news interests users. We use this information to understand to what extent the sentiment policy employed by news media have been successful in catching users' attention. We show that negative news receive more number of comments and shares compared to positive news, which gets more number of likes compared to negative news. This finding is extremely interesting as it agrees with the popular `negativity bias’ theory\footnote{A popular theory in social psychology that states that humans are more likely to focus on bad news~\cite{rozin2001negativity}.} for actions that require greater involvement such as sharing and commenting but does not agree for the relatively simpler actions such as liking the news article.

Comments allow users to express their opinion regarding a news item. These opinion can be used for opinion mining to gather information on how users perceive the news, predict real-world outcomes, gain useful insight into users’ collective behavior, etc. These mining tasks often involve aggregating the users’ opinions from different news channels, which may potentially bias the result because users' opinion on a topic depends on many factors~\cite{cambria2013new,vinodhini2012sentiment}, such as the region where the news is published, the time when it is published, the type of information source that published the news, the sentiment with which the news is written, etc. In this paper, we analyze how users' opinion depend on two of these factors: sentiment of the news article and type of news channel. We obtain interesting insights that can be used to correct the bias arising due to the wavering nature of users' comments. To the best of our knowledge, very few studies~\cite{bautin2008international,godbole2007large} have been devoted to the sentiment analysis of news articles and none has arrived to study the role of information sources coupled with the sentiment of news on the users’ perception of the news. To gain more insight into the factors affecting the sentiment of news, we also categorize the news based on the topic, time, and their significance. 

Some of our major findings show that sentiment generated by social media channels of different types of information sources are different and most of the time, these news channels generate either positive or negative news. Surprisingly, print and radio based channels generate predominantly positive news on their social media pages disagreeing with the popular opinion~\cite{patterson2011out,stieglitz2013emotions} that news sites mostly post negative news to take advantage of the negativity bias. We also found that negative news were shared and commented on more often but positive news were liked more often throwing light on how the negativity bias operates at a different level of user engagement. Additionally, we show that polarity of the comments is strongly related to the sentiment polarity of the news. As news become more negative, their comments also become more negative in tone and vice versa. News from TV based news channels prompt more negative reactions from news compared to print and radio based channels, suggesting that people react not only to the type of news article but also to the source of the news article.

\vspace{0.1in}
\noindent Our key contributions are as follows: 
\begin{itemize}
\item We analyze sentiment of posts created by news channels in their social media pages. 
\item We investigate users' reactions to news posts of varying sentiment from different types of news channels. 
\item We categorize the news posts into different topics to gain insight into the sentiment of news posts created under these topics. 
\item We compare the sentiment of big news headlines with niche news to investigate how big headlines impact sentiments of the news posts. 
\item We explicate the relationship between sentiment polarity of news articles and the polarity of textual reactions in conjunction with type of information source. 
\item We perform temporal analysis to investigate the posting behavior of news articles having different sentiments over a period of time. 
\item We perform our experiments on a large dataset containing 1.13 billion users reactions and 0.15 million news articles. 
\end{itemize}

The rest of the paper is organized as follows. In Section~\ref{sec:sa_rw}, we briefly survey the related work. 
Section~\ref{sec:sa_meth} presents our methodologies. 
Section~\ref{sec:polnp} analyzes the polarity of different types of social media news channels. Section~\ref{sec:popvspol} describes the relationship between the popularity and polarity of news articles. In section~\ref{sec:anuo}, we analyze the users' opinions on different types of channels with varying post sentiments. We present temporal analysis in Section~\ref{sec:senti_tempana} and conclude our work in Section~\ref{sec:sa_cnfw}.

\section{Related Work}
\label{sec:sa_rw}

Online news content analysis is an active research topic in Computer Science. One of the major sub-topic is to predict the popularity of news content~\cite{ahmed2013peek,bandari2012pulse,lee2010approach,lerman2010using,szabo2010predicting,tatar2014popularity}.  Lerman et al.~\cite{lerman2010using} estimated the overall popularity of a content based on early users' reactions on it.  However, Bandari et al.~\cite{bandari2012pulse} predicted the popularity of a content without using early users' reactions. They considered multiple features of a news post to predict its popularity prior to its posting, such as subjectivity of the news, category of the news, etc. In this paper, we study the relationship between news post sentiment and the news post popularity across different types of news channels. Unlike popularity prediction studies, we analyze the sentiment of news articles, and it’s effect on popularity and users’ opinions.

Several approaches~\cite{arapakis2014user,coscia2014average,hansen2011good,naveed2011bad,wu2011does} have been proposed to analyze the news propagation. Naveed et al.~\cite{naveed2011bad} showed that negative news is more attractive to users and easily catch their attention. They have used 15 different set of content-based features and predict the likelihood of a tweet being retweeted using logistic regression. 
Wu et al.~\cite{wu2011does} showed that the lifetime of negative news is very short but positive news stay for a longer time. They predict the decay of social media content using a classification technique namely, support vector machine. Coscia et al.~\cite{coscia2014average} studied the content of news and showed that news content should be significantly different compared to already existing content, as unique content spreads faster than average or existing content. However, we show that it is not only the negative news that catch user attention but also positive news can garner a lot of user attention. Popularity gained by both positive and negative news is usually higher than neutral news.

Recently, researchers~\cite{reis2014magnet} at UFMG developed a tool that present news to users based on their interest or polarity. They ranked the news articles based on their popularity and sentiment score. Reis et al.~\cite{dos2015breaking} analyzed the news headlines from a popular global news channel. They showed that sentiment of headline correlates with the popularity of news and negative comments are posted independently of the sentiment score of the headlines. Diakopoulos and Naaman~\cite{diakopoulos2011topicality} analyzed the relationships between news comment topicality, temporality, sentiment, and quality. They showed that comment sentiments (positive and negative) are the important indicators of discourse quality. Our analysis is complementary to existing studies, which shows that polarity of comment is not completely independent of polarity actual post; it is a function of the polarity of the news post.

Further, a work published on the social news on the web\footnote{http://snow2013.isti.cnr.it/?cat=4}, dealt with the problem of finding the topics of a news channel and users’ interests in social media~\cite{zubiaga2013newspaper}. Authors of the paper performed analysis on The New York Times channel and showed that news stories selected by editors for their newspaper or channel are far from those catching users’ attention on social media. However, in this paper, we perform sentiment analysis on multiple social media news channels and show the top topics of editor interests as well as sentiment associated with each topic. Unlike the findings reported by Zubiaga~\cite{zubiaga2013newspaper} that news channels mainly report hard news with top priority, we have found that news channels uniformly generate hard news (e.g., politics, money, world) as well as soft news (e.g., entertainment, lifestyle, sports, science and technology) through their social media pages. 
Another study by Nies et al.~\cite{de2013towards} showed the impact of an actual news article on social media. They compared the relevancy of an arbitrary news article with social media news to measure its impact. However, we analyze the impact of social media news content with the varying sentiment on social media users. We show how a news with different sentiments shapes the users’ opinion about the news content.

Compared to existing works, our focus is on analyzing the sentiments of news posts in social media and users’ reactions to it. We examine the correlation between the polarity of comments and the sentiment polarity of posts. Unlike previous studies~\cite{burke2016once,castillo2014characterizing,cheng2014community,d2013dont,moosa2014comment}, our analysis shows that the sentiment of comment is a function of the news post sentiment and the type of channel through which news was traditionally disseminated.

\section{Methodology}
\label{sec:sa_meth}
In this section, we first describe the process of collecting the news from Facebook pages of news channels. We then describe the method employed to measure the sentiment polarity. Next, we present the method to categorize the news posts. In the rest of the paper, we use terms such as ‘post’, ‘news post’, ‘news content’, and ‘news article’ interchangeably.

\subsection{News Posts Collection}
In order to characterize the news posted on social media, we collected the news from Facebook pages  of five major news media channels. Our choice of Facebook over other social networking platforms is informed by research from Pew Research center, which notes that Facebook has the highest reach with 44\% of adults in the US getting news on the platform~\cite{gottfried2016news}. Further, we chose news sites with the highest valuation as calculated by Virtue’s Social Page Evaluator\footnote{http://www.adamsherk.com/social-media/most-valuable-news-site-facebook-pages/}. In order to understand the differences between different media, we choose two each of television and print media based channels and one radio based channel. Dataset thus includes posts from the Facebook pages of CNN and Fox News which are television news channels, The Economist and The New York Times (or NYT) which are daily and weekly newspaper organizations respectively, and NPR which is a public radio network. 

We extract the dataset\footnote{The dataset will be made available for download from the author's website.} from Facebook pages using the Facebook Graph API~\cite{facebook:graph}. The dataset contains news articles posted by the pages, reactions on the post, link to the original news article and attributes including the number of users who liked the page, organization name, post creation time, reaction time, etc. Users can react to posts created by pages in the form of like, comment and share. Reactions consist of textual comments and rating score in the form of likes and shares. 
For each news channel, we present the number of posts, comments, likes, shares, and time interval in the collected news dataset as follows:

\begin{table}[!h]
\begin{center}
 \small
  \begin{tabular}{|l|l|l|l|l|l|}
\hline
\hline
 \thead{News Channels} & \thead{Posts} & \thead{{Comments}}  & \thead{{Likes}}  & \thead{{Shares}}  & \thead{{Time Interval}}\\
 \hline\hline
CNN &33324 & 26582081 &147310056 & 52936764 &  Dec 2016-April 2012\\
 \hline
NPR & 18266 & 4585776 & 56007054 & 18847580 & Dec 2016-Nov 2013\\
 \hline
Fox News & 26525 & 83957661 & 443933576 & 143762565 &  Dec 2016-Jan 2014\\
 \hline
The Economist & 24272 & 1336956 & 20206137 & 6376644 & Dec 2016-Dec 2014\\
 \hline
NYT  & 47522 & 9226029 & 93891025 & 25616593 &  Dec 2016-April 2013\\
 \hline
\end{tabular}
 \caption{Dataset Statistics}
\label{table:sa_ds}
\end{center}
\end{table}

From the Table~\ref{table:sa_ds}, we can infer that Fox News is the most popular news channel as it has the highest reaction per post ratio. Whereas The Economist is the least popular news channel among all the news channels as it has the lowest reaction per post ratio. Here, the reaction is the popularity measure which is the sum of likes, comments, and shares. We also perform pre-processing to remove noisy and unimportant words from the textual posts and comments. It avoids the trivial words, which appear frequently in the posts. We remove stop-words, such as `a', `an', `the', etc., as these words do not contain significant information for our analysis. We also employ stemming and lemmatization~\cite{uysal2014impact} to reduce inflected or derived words to their root forms. Throughout the rest of the paper, we consider a common time frame from December 2014 to December 2016 for our analysis.

\subsection{Sentiment Polarity Identification}

In social media, users use informal language to present their textual contents which differentiate social media texts different from standard texts. We are listing a few examples of the usage of informal language as follows: \begin{itemize}
\item  Social media texts especially comments usually contain emoticons such as :),  :(,  :-), |-o, etc. 

\item  Increasing number of users use acronyms such as LOL, smh, ty, wth, etc. 

\item  Social media users use slang words very often, and these words became a part of social media lexicon. For example, meh, yep, giggly, nah are few commonly used slang words. 

\item  Users also use multiple punctuation marks to emphasize the certain words in a text sentence. 
\end{itemize}

In order to tackle all the above-mentioned issues, we have used Valence Aware Dictionary and sEntiment Reasoner (VADER), which is a powerful sentiment analyzer to find the sentiment of social media texts~\cite{hutto2014vader}. VADER is a lexicon and rule-based sentiment reasoner and is the best suited for sentiment analysis of contents originating in social media~\cite{tamersoy2015characterizing}. It creates and utilizes a new gold standard sentiment lexicon with 7500 lexical features that are commonly used to express the sentiment in a social media text. It uses the rule based method consisting of five rules that embody grammatical and syntactical conventions for expressing the sentiment intensity\footnote{https://github.com/cjhutto/vaderSentiment}. VADER has been compared with 11 sentiment analysis tools/techniques including SentiWordNet~\cite{baccianella2010sentiwordnet}, SenticNet~\cite{cambria2012senticnet}, LIWC~\cite{pennebaker2001linguistic}. It is shown that VADER outperforms all of them. Further, VADER provides a sentiment score in the range of -1 to +1, with -1 being extremely negative, +1 being extremely positive and, 0 being neutral. For the sake of interpretability, we convert these polarity scores to an integer between -5 to +5. The polarity scores inferred were as expected and a sample of the same can be observed in Table~\ref{table:sa_spsp}.

\begin{table}[!h]
\begin{center}
\resizebox{\columnwidth}{!}{%
\small
\begin{tabular}{|p{1cm}|p{14cm}|}
\hline
\hline
{Score} & {Sample Post}\\
 \hline\hline
+5 & It’s just an amazing thing to watch good old-fashioned regular human beings and a whole lot of love change the world seismically\\
\hline
+4 & Follow the Queen’s Diamond Jubilee celebrations with the latest photos, videos, facts and trivia. Tell us which part of the festivities you’re most impressed with\\
\hline
+3 & When you do something extraordinary, it's shown that you can inspire other people." \#CNNHeroes\\
\hline
+2 & The world's first permanent ice hotel has opened in Sweden, thanks to new solar-powered cooling technology\\
\hline
+1 & Farmers in the Australian desert are growing 15,000 tons of tomatoes using seawater — and thousands of mirrors\\
\hline
 0 & In a tweet, President-elect Donald J. Trump says his businesses won't do any new deals while he's in office\\
\hline
-1 & Between 2007 and 2014, 30\% of African elephants disappeared\\
\hline
-2 & Being exposed to the daily hassles of traffic can lead to higher chronic stress and higher blood pressure,” according to a recent study conducted in Texas\\
\hline
-3 & Are we on the verge of a second Cold War?\\
\hline
-4 & Terror attacks have ripped apart small towns and big cities across the Middle East and Africa throughout 2016, and this weekend was no different\\
\hline
-5 & A young newlywed couple died a horrible death at the hands of the bride's family\\
\hline
\end{tabular}%
}
\caption{Sentiment polarity of sample posts}
\label{table:sa_spsp}
\end{center}
\end{table}

\subsection{News Posts Categorization}
To get insight into posting behaviour of news channels across the categories, it is useful to categorize the news posts into multiple categories such as sport, entertainment, politics, science and technology (sci\&tech), etc. 
Unlike online news sites, news posted on social media channels is not categorized. In order to categorize these news posts, we use unsupervised method LDA~\cite{blei2003latent}. LDA is a probabilistic topic modeling algorithm which represents each document (in this case, a news post) as a mixture of various topics with definite probabilities ($\theta$). A topic is comprised of words or terms. The terms that often occur together, are placed under the same topic with high probabilities ($\phi$). Document-topic distribution ($\theta$) and term-topic ($\phi$) are computed using Gibbs sampling as follows:

\begin{equation}
\theta_{dj} = \dfrac {C^{DT}_{dj} + \alpha}{\sum_{k=1}^T C^{DT}_{dk} + T\alpha}
\label{eq:1}
\end{equation}

\begin{equation}
\phi_{ij} = \dfrac {C^{WT}_{ij} + \beta}{\sum_{k=1}^W C^{WT}_{kj} + W\beta}
\label{eq:2}
\end{equation}

where $T$, $D$ and $\alpha$ represent the number of topics, documents, and smoothing constant respectively. $C^{DT}_{dj}$ is the number of times a term appears in document $d$ that has been assigned to topic $j$. $W$,  $T$ and $\beta$ represent the number of terms, topics, and smoothing constant respectively. $C^{WT}_{ij}$ is the number of occurrences of a word $i$ that has been assigned to topic $j$. Gibbs sampling method integrate these two assignments and update the topic assignment until convergence. 

In order to make the topic modeling richer, we augment the post message with its URL information obtained using the Graph API. The augmented text from external documents using URL was later processed to remove invalid characters and corrected for mistakes in spelling. 
To determine the ideal number of topics $k$, we perform 5-fold cross validation on perplexity at different values of $k$. We then compute the rate of perplexity change (RPC)~\cite{zhao2015heuristic} on a 10\% random sample. Perplexity is a statistical measure and often used to measure the performance of topic models~\cite{Griffiths5228}. Perplexity reflects the capacity of a model to generalize to test set or unseen posts. The point where the rate of perplexity no longer falls significantly with an increase in the number of topics is used as the ideal number of topics. In our experiment, we find the optimum value of $k$ is 10 from where perplexity does not change significantly.

As studied by Chang et al.~\cite{chang2009reading} that perplexity and human judgment are not well correlated, we evaluate our topics manually using precision~\cite{agrawal2009diversifying}. We ask five research scholars having knowledge of topic modeling to judge the relevancy of topical words generated by the topic model. We ask research scholars to label each topical word as relevant or non-relevant to assigned topic by the topic model. Topics are labeled by researchers independently without influencing each other. Topics for which researchers did not agree on were discussed until a consensus was reached. We then compute precision as a fraction of generated topical words that are relevant to the assigned topic. We find that the topic model performs reasonably well with 80.3\% precision. One of the reasons for this is that number of topics selected for LDA categorization is the best suited for Facebook news posts dataset. Moreover, posts created by news channels in their social media pages are well framed unlike user-generated contents such as comments, tweets, etc.

Further, we provide the label for each topic based on the most relevant terms that uniquely define the topic. Since each topic contains thousands of terms, we extract top relevant words based on term-topic distribution. Relevance ($r$) of a term $i$ to topic $j$ is computed as follows: 

\begin{equation}
r_{ij} = \lambda\log(\phi_{ij}) + (1-\lambda)\log(\frac{\phi_{ij}}{p_i})
\label{eq:3}
\end{equation}

where $\phi_{ij}$ is the term-topic probability, and $p_i$ is the empirical probability of the word in the corpus. $\lambda$ is a weighting term and we choose 0.6 as an optimal value for $\lambda$ as shown by  Sievert et al.~\cite{sievert2014ldavis}. We assign each post or document to these labeled topics based on document topic probability ($\theta_{dj}$). If document $d$ shows the highest probability for topic $j$, $d$ is assigned to topic $j$. 

\begin{figure}[H]
 \centering
  \includegraphics[width=11cm, height=6.5cm]{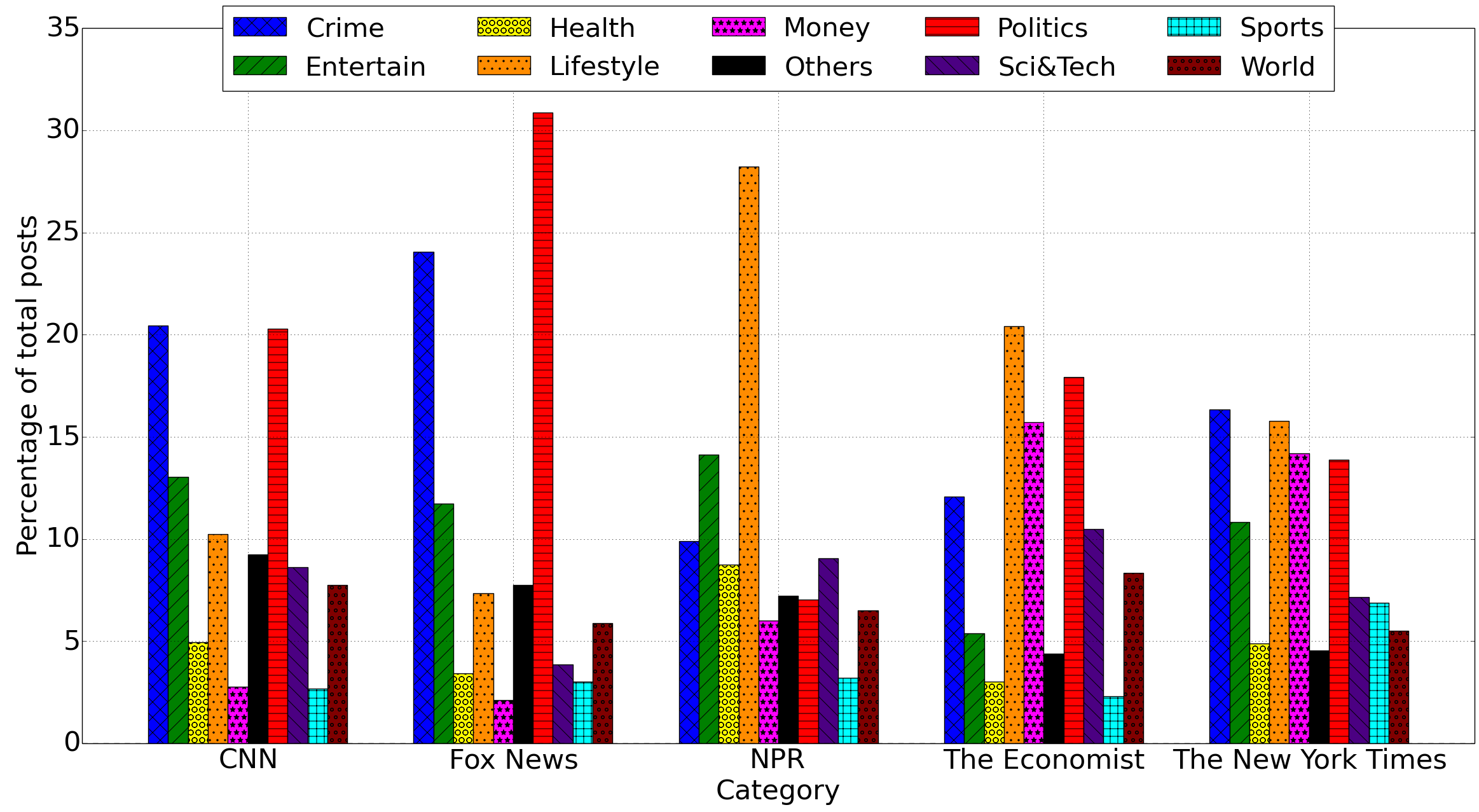}
  \caption{Distribution of news posts across categories}
  \label{fig:all_cat}
\end{figure}

Figure~\ref{fig:all_cat} shows the distribution of posts for each channel across categories. We observe that post distribution of TV based channels CNN, Fox News is almost similar where politics, crime categories contain a higher percentage of posts. In case of print media based channels like The New York Times and The Economist, lifestyle, money, politics, and crime news seem to be more common. On the other hand, NPR which is a Radio based channel, posts lifestyle news most often followed by entertainment (entertain). We show how channels post the news in these categories in Section~\ref{sec:polcat}.

\section{Analysis of News Posts Polarity}
\label{sec:polnp}

We begin our investigation by analyzing the distribution of polarity of post messages grouped as positive, negative and neutral for the social media news channels as discussed in Section~\ref{sec:sa_int}.

\begin{figure}[!h]
    \centering
  \includegraphics[width=10cm, height=5cm]{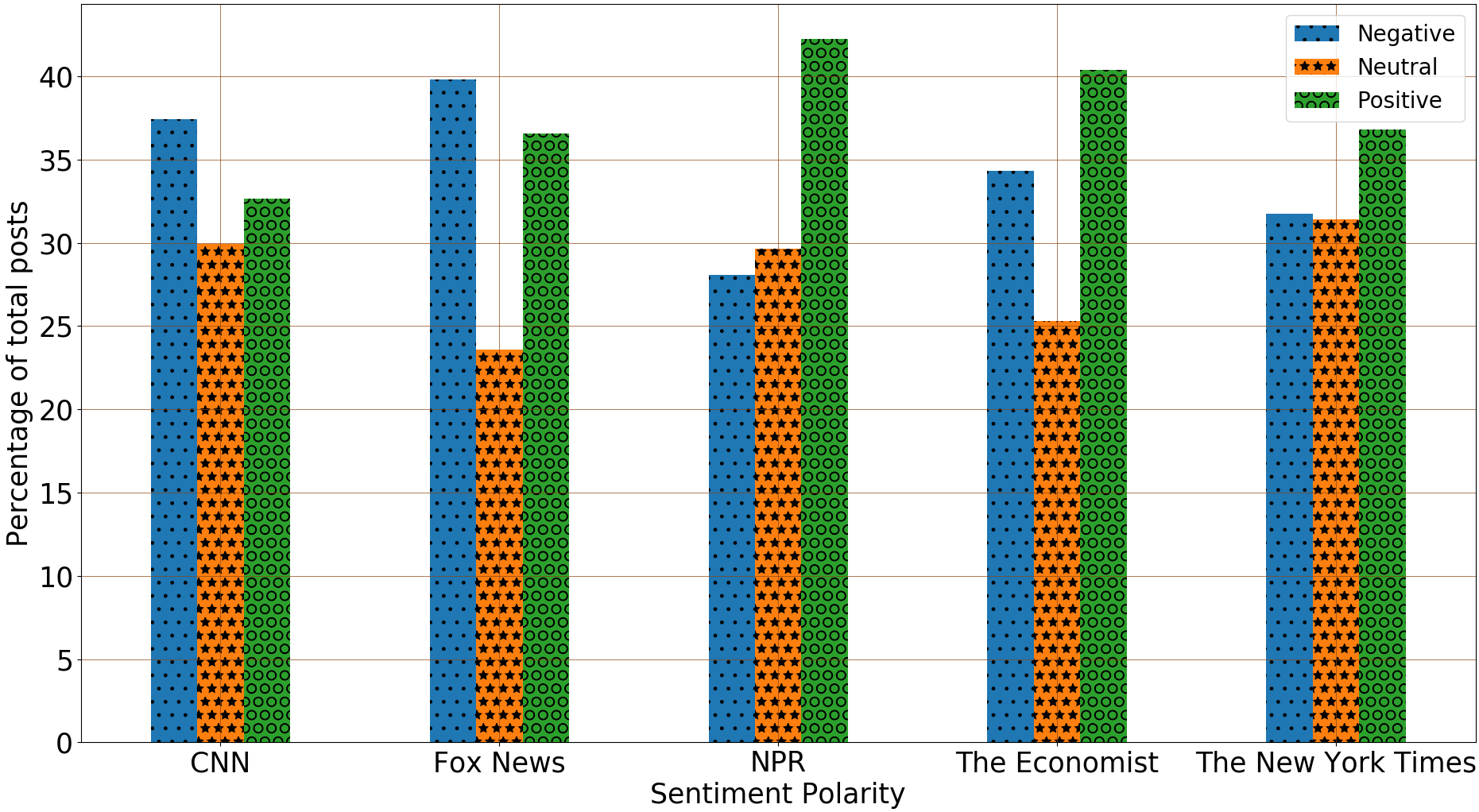}
    \caption{ Polarity of news posts generated by pages}
    \label{fig:sa_polar}
\end{figure}

A quick glance at Figure~\ref{fig:sa_polar} reveals that the dominant sentiment in posts by all the news channels is always either positive or negative but not neutral. Moreover, posts with neutral sentiment are least common in Facebook pages of all media sites except NPR. These inferences support the claim that all the channels tend to generate more positive or negative messages on their Facebook pages to attract the users' attention.

Another interesting aspect to be noted is the similarity in the distribution of sentiment polarity of posts between media channels that function through the same medium of communication. 
Posts by television based news channels, such as Fox News and CNN, are predominantly negative where Fox News generates the highest percentage (40\%) of negative news across all the channels. On the other hand posts by radio and print media based channels such as NPR, The Economist and The New York Times are mostly positive. Radio based news channel, NPR generates the highest percentage (43\%) of positive news and the least percentage of negative news (28\%) across all the channels. 
Print based media channels, The Economist and The New York Times generate a large proportion of positive news and a less proportion of neutral news. Despite the similar pattern of news generation by these two channels, The Economist generates relatively a larger percentage of both the positive and negative news compared to the neutral news. One of the reasons for this is that The Economist reports growth (i.e. positive news) and decline (i.e. negative news) in business, commerce, and trade substantially. 
To investigate how different types of channels report the same news, we have also performed the sentiment experiment on the same news events covered by these channels. However, we did not notice a significant difference in the sentiment compared to Figure~\ref{fig:sa_polar}. We have observed the sentiment pattern similar to Figure~\ref{fig:sa_polar} for all the channels.


The news reported by news sources has evolved differently because of the manner in
which users consume the information in each medium~\cite{bennett2016news}. Our analysis suggests that these differences remain despite disseminating information on a common platform. News media are often criticized for their focus on negative news rather than providing a balanced picture of the world~\cite{hansen2011good,budak2016fair,leetaru2011culturomics}. This phenomenon has been attributed to journalistic cynicism and inherent preference for negative news among users. However, we observe through our analysis, that print and radio based social media channels post more positive news than negative news. This finding raises important questions: Is this change in the type of content posted by print and radio based channels precipitated by user’s preference for positive news on social networking platforms? Does this mean that the popular negativity bias theories~\cite{rozin2001negativity,baumeister2001bad}, which state that humans have a predilection for negativity, not hold true in the case of news consumption in social media? We attempt to answer these questions in Section~\ref{sec:popvspol}. 

\subsection{News Posts Polarity across Categories}
\label{sec:polcat}
In this section, we analyze the polarity of news posts across categories to investigate how news channels generate the news across categories. We compare the polarity of news generated in multiple categories such sports, politics, health, entertainment, etc. We observe that channels from the similar type of sources show the similar pattern. Due to space constraints, we present the results of only one news channel of each type of information sources such as print, television, and radio as follows:

\begin{figure}[!h]
\minipage{0.5\textwidth}
  \includegraphics[width=7.5cm, height=5cm]{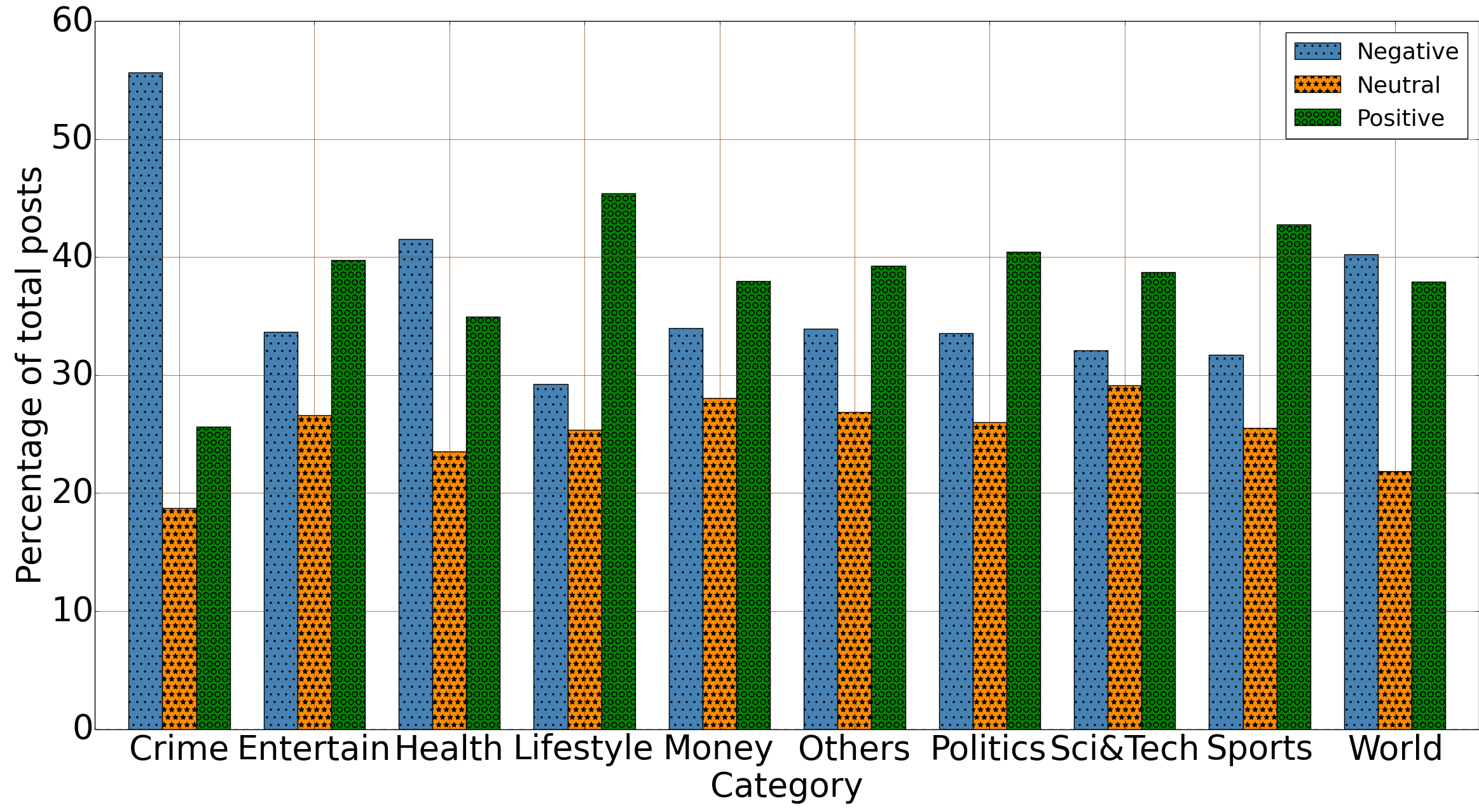}
  \caption{Fox News}
  \label{fig:fox}
\endminipage\hfill
\minipage{0.5\textwidth}%
  \includegraphics[width=7.5cm, height=5cm]{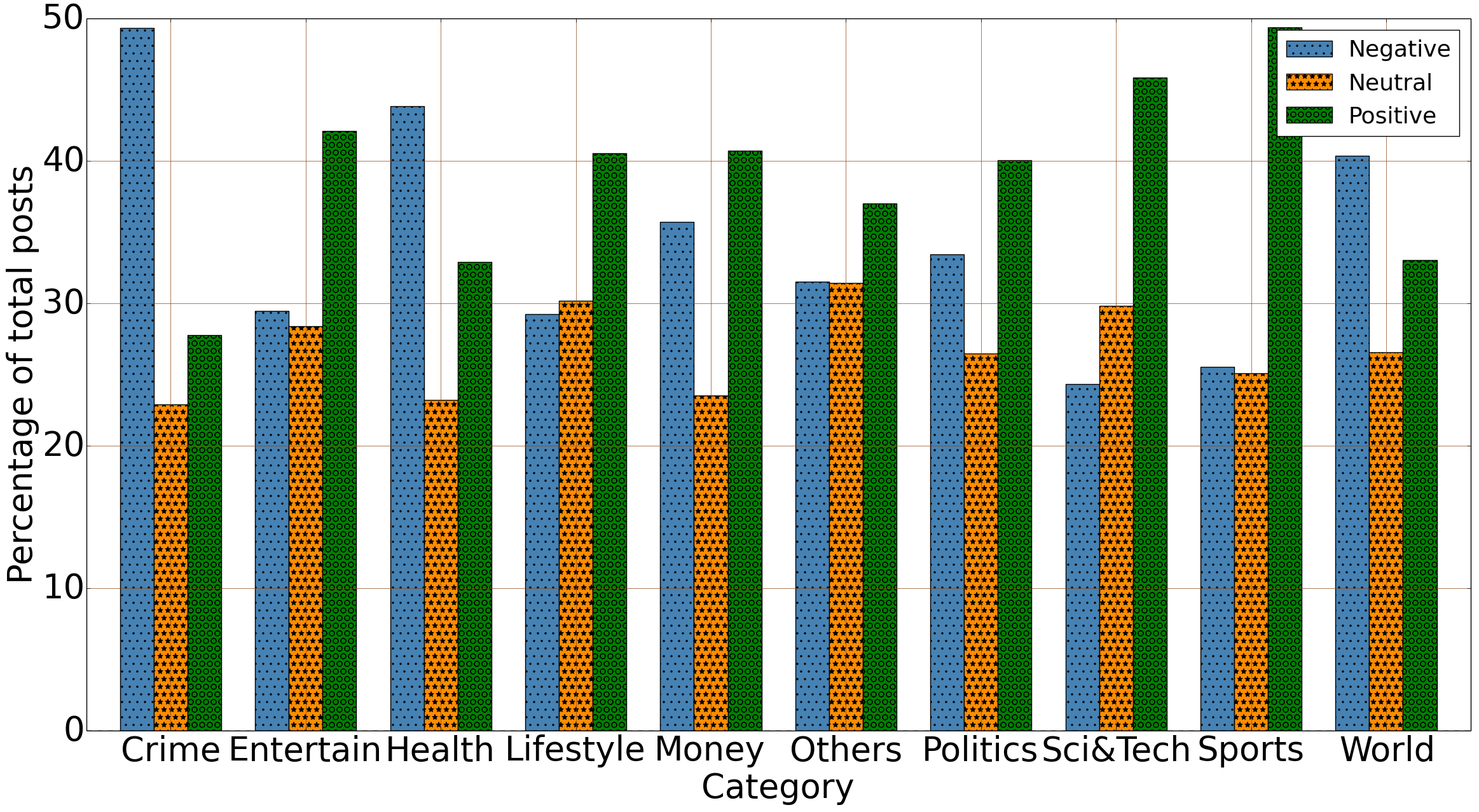}
  \caption{The Economist}
  \label{fig:econ}
\endminipage
\end{figure}

\begin{figure}[!h]
\centering
\includegraphics[width=10cm, height=5cm]{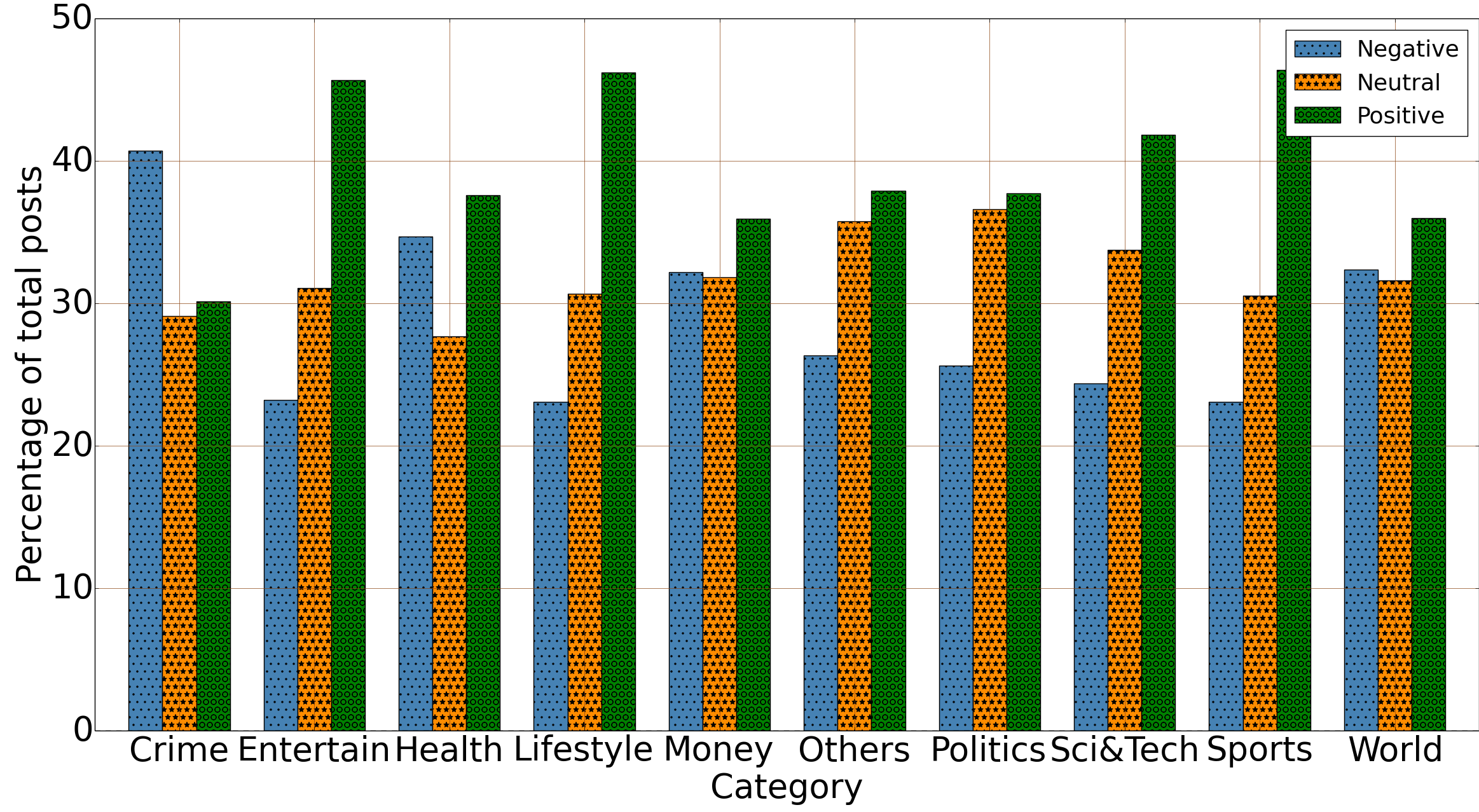}
  \caption{NPR}
  \label{fig:npr}
\end{figure}

It can be observed from Figures~\ref{fig:fox} and \ref{fig:econ} that news belonging to the crime, world and health categories are predominantly negative, for both print and television based channels. 
One of the reasons for this is that most of the times news related to crime is woeful and unpleasant. The news related to health and world easily catch the attention of the channels if any terrible accident takes place across the world.  
However, in case of NPR (Figure~\ref{fig:npr}), which is a radio based channel, all types of news except crime news, are predominantly positive in tone. This suggests the possibility that the trend in dominant sentiment observed in the Section~\ref{sec:polnp} could also be a result of the same type of news being covered differently by different news channels depending on the primary medium. Thus, by analyzing the sentiments across categories for different organizations, we can conclude that both the differences in the type of news that is often covered, and the difference in the tone with which the same news is covered are responsible for the difference in dominant sentiment observed in the previous section. 

Moreover, except for NPR, across all categories, the proportion of news that is neutral in tone is the smallest. It indicates that trend of a higher fraction of positive or negative news, which we observed in the Section~\ref{sec:polnp}, is not limited to a few categories but is one of the tactic that is adapted for the generation of all types of news. NPR, however, stands out with negative news being the least common in majority of the categories. 
This observation is also consistent with the predominantly positive nature of news generated by NPR that we observed in the previous section. 
Moreover, the similarity in the sentiment distribution for news channels using the same medium further asserts the influence of the medium on the tone with which news is disseminated by channels.

\subsection{Big headlines versus niche news}
\label{sec:headlinevsniche}

In this section, we compare the polarity of big headlines and niche news. We analyze the differences in the sentiment of news posts reported by different types of channels for big headlines as well as niche news. We report the findings in the following figures:

\begin{figure}[!h]
\minipage{0.5\textwidth}
  \includegraphics[width=7.5cm, height=5cm]{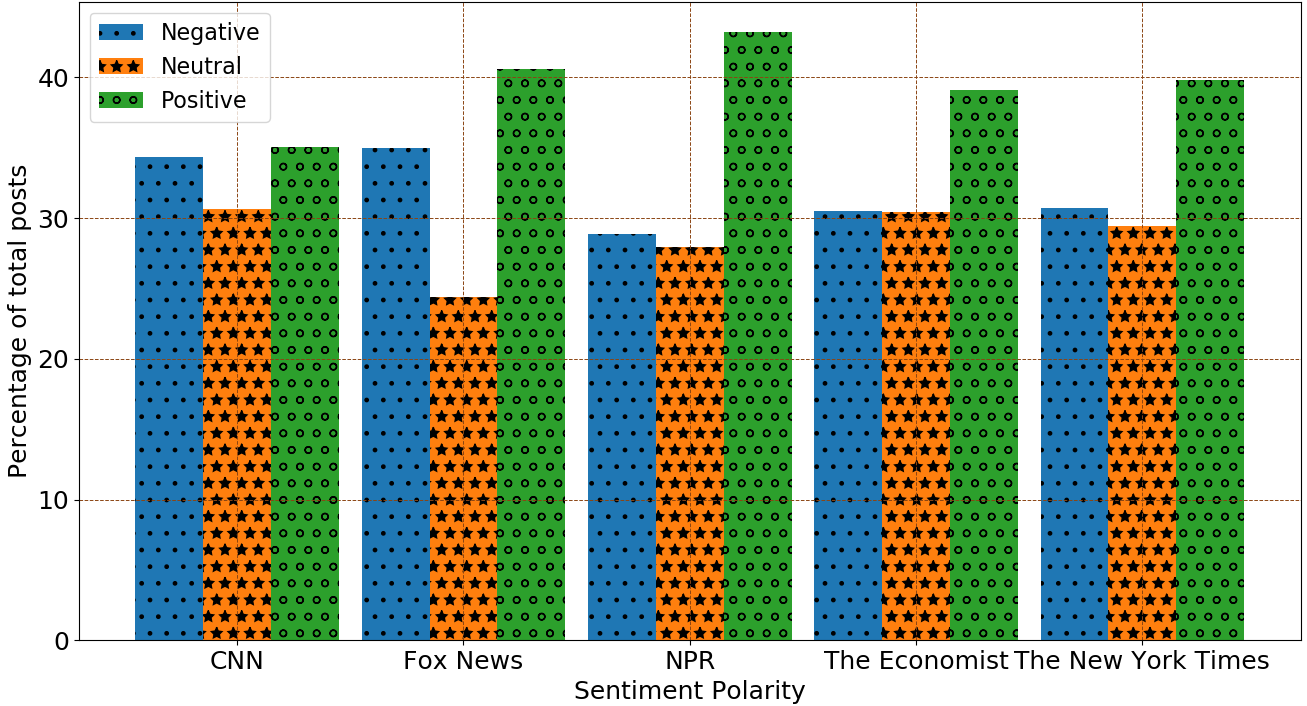}
  \caption{Big headlines}
  \label{fig:headlines}
\endminipage\hfill
\minipage{0.5\textwidth}%
  \includegraphics[width=7.5cm, height=5cm]{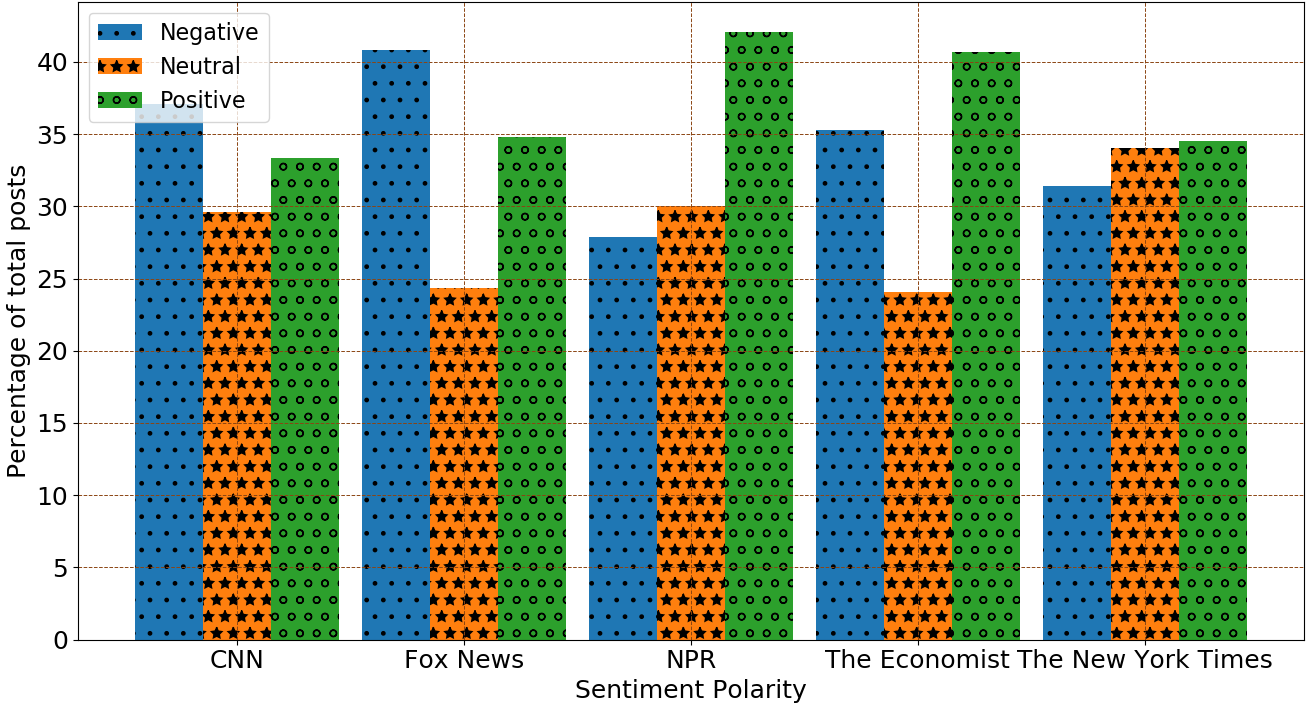}
  \caption{Niche news}
  \label{fig:nichenews}
\endminipage
\end{figure}

As can be seen in Figures~\ref{fig:headlines} and \ref{fig:nichenews}, the sentiment polarity of big headlines are different from the polarity of niche news. All the channels generate a higher percentage of positive headline news compared to negative and neutral news (refer to Figure~\ref{fig:headlines}). Among all the channels, NPR generates the highest percentage (43\%) of the positive headline news. One of the reasons for generating a higher percentage of positive news is that these big headlines are very popular and exist for a longer time. If news channels continuously generate a higher fraction of negative news for these types of events, users may lose their interests and it would lead to less engagement on these news channels. Negative news goes away very fast~\cite{wu2011does} and if there is a big headline that is usually persistent for some time, channels create more number of positive news to maintain the sustainability. As studied by researchers in psychology~\cite{johnston1997psychological} that negative news causes worries to users, channels generate more positive news about the headline to retain the users’ interests over time. On the other hand, we do not observe a significant difference in the polarity of niche news as compared to Figure~\ref{fig:sa_polar}. Positive or negative news are more popular than the neutral news. TV  based channels report more negative news compared to the radio and print media based channels whereas print and radio based channels report more positive news.

\subsection{Polarity of Same News Events across Channels}
In this section, we examine how different types of news channels report the same news. We analyze sentiments of ten different real-world news events that are posted by social media news channels. In Table~\ref{table:polrsamenews}, positive indicates the percentage of positive news created for the event and negative indicates the percentage of negative news created for the event. Due to brevity, we did not mention the neutral news, which can be determined by subtracting the sum of positive and negative news from hundred.

\begin{table}[!h]
\begin{center}
\begin{center}
\begin{tabular}{|l|l|l|l|l|l|}

    \hline\hline
   \thead{ News Channel / \\ News} &  \thead{CNN} & \thead{Fox News} & \thead{NPR} & \thead{NYT} & \thead{The Economist}\\
    \hline \hline
     
\thead{Presidential Election} &  \thead{ Positive: 42.3\% \\ Negative: 22.5\%}  &  \thead{ Positive: 61.5\% \\ Negative: 18.2\%}  &  \thead{ Positive: 55.3\% \\ Negative: 8.1\% } &  \thead{ Positive: 48.7\% \\ Negative: 18.7\%}  &  \thead{ Positive: 41.9\% \\ Negative: 23.8\%} \\
    \hline

\thead{Same-sex Marriage} &  \thead{Positive: 45.6\% \\ Negative: 32.8\%}  &  \thead{Positive: 52.3\% \\ Negative: 40.7\%}  &  \thead{Positive: 45.2\% \\ Negative: 40.4\% } &  \thead{Positive: 60.5\% \\ Negative: 19.8\%}  &  \thead{Positive: 65.3\% \\ Negative: 18.6\%} \\
    \hline

\thead{Obamacare} &  \thead{Positive: 50.3\% \\ Negative: 32.5\%}  &  \thead{Positive: 42.1\% \\ Negative: 30.6\%}  &  \thead{Positive: 45.2\% \\ Negative: 21.7\% } &  \thead{Positive: 41.3\% \\ Negative: 35.6\%}  &  \thead{Positive: 65.2\% \\ Negative: 16.8\%} \\
    \hline
\thead{Football} &  \thead{Positive: 35.7\% \\ Negative: 40.5\%}  &  \thead{Positive: 38.9\% \\ Negative: 42.4\%}  &  \thead{Positive: 45.2\% \\ Negative: 18.7\% } &  \thead{Positive: 45.3\% \\ Negative: 35.6\%}  &  \thead{Positive: 52.3\% \\ Negative: 31.6\%} \\
    \hline
 \thead{Dakota Access Pipeline} &  \thead{Positive: 37.4\% \\ Negative:42.3\%}  &  \thead{Positive: 38.5\% \\ Negative:40.9\%}  &  \thead{Positive: 60.4\% \\ Negative: 25.7\% } &  \thead{Positive: 39.4\% \\ Negative: 38.5\%}  &  \thead{Positive: 50.6\% \\ Negative: 28.3\%} \\
       \hline
 \thead{US Ambassador} &  \thead{Positive: 32.3\% \\ Negative:45.7\%}  &  \thead{Positive: 37.7\% \\ Negative:48.6\%}  &  \thead{Positive: 35.3\% \\ Negative: 25.6\% } &  \thead{Positive: 48.5\% \\ Negative: 41.3\%}  &  \thead{Positive: 41.4\% \\ Negative: 35.4\%} \\
    \hline
 \thead{Hollywood} &  \thead{Positive: 32.4\% \\ Negative:42.8\%}  &  \thead{Positive: 35.2\% \\ Negative:43.5\%}  &  \thead{Positive: 43.6\% \\ Negative: 27.5\% } &  \thead{Positive: 45.2\% \\ Negative: 30.3\%}  &  \thead{Positive: 53.6\% \\ Negative: 25.7\%} \\
    \hline
 \thead{ Ebola} &  \thead{Positive: 39.4\% \\ Negative:42.3\%}  &  \thead{Positive: 34.7\% \\ Negative:39.8\%}  &  \thead{Positive: 38.6\% \\ Negative: 46.5\% } &  \thead{Positive: 28.3\% \\ Negative: 45.7\%}  &  \thead{Positive: 25.4\% \\ Negative: 56.8\%} \\
    \hline
\thead{MH370 Flight} &  \thead{Positive: 15.3\% \\ Negative:50.2\%}  &  \thead{Positive: 10.3\% \\ Negative:60.4\%}  &  \thead{Positive: 6.8\% \\ Negative: 75.6\% } &  \thead{Positive: 8.4\% \\ Negative: 60.3\%}  &  \thead{Positive: 5.3\% \\ Negative: 66.8\%} \\
    \hline
\thead{Zika Virus} &  \thead{Positive: 19.4\% \\ Negative: 41.6\%}  &  \thead{Positive: 34.2\% \\ Negative: 47.5\%}  &  \thead{Positive: 28.8\% \\ Negative: 50.3\% } &  \thead{Positive: 10.4\% \\ Negative: 48.8\%}  &  \thead{Positive: 31.5\% \\ Negative: 40.3\%} \\
    \hline
\end{tabular}%
\end{center}
\caption{Sentiment polarity of news events across channels}
\label{table:polrsamenews}
\end{center}
\end{table}

As can be seen in Table~\ref{table:polrsamenews}, different news channels report the same news events differently. Despite being generating the same news with different percentage of sentiment polarities, all the channels generate a large fraction of positive news for big headlines such as Presidential election, Same-sex Marriage, and Obamacare (refer Section~\ref{sec:headlinevsniche}). However, if a headline is very negative in nature such as MH370 Flight Disappearance, all the news channel generate a large fraction (more than 50\%) of negative news due to nature of the news event. Also, all the channels generate a major fraction of negative news, if a news is related to flu epidemics such as Ebola and Zika Virus, an attack, and natural disaster. 

Although all the big-headlines are reported with a similar pattern of sentiment, regular events and the events that are not part of big-headlines are reported differently by the channels. If a news is related to regular events or minor headlines (i.e., Football, US Ambassador, Hollywood, Dakota Access Pipeline), channels usually generate a different sentiment pattern of positive, negative and neutral news. In this case, positive or negative news are more popular than the neutral news. TV  based channels report more negative news compared to the radio and print media based channels. On the other hand, print and radio based channels report more positive news. This observation is in line with our analysis in Section~\ref{sec:polnp} and due to the majority of these regular events, we observe the similar pattern of sentiment in Figure~\ref{fig:sa_polar}. 

Apart from the results that are shown in Table~\ref{table:polrsamenews}, we also observe that news related to major breakthrough in Science and Technology (i.e., news from NASA, MIT), highly reputed awards (i.e., Nobel Prize, Oscars) and persons (i.e., Pope, Dalai Lama) are reported significantly positive (more than 55\%) across all the channels. One of the reasons is that these are very esteemed organizations, awards or persons. Due to the highly positive nature, all the channels generate a major fraction of the positive news related to these organizations, awards, and persons.

\section{Popularity versus Polarity}
\label{sec:popvspol}
We analyze the popularity of a news post as a function of its polarity. Affinity metrics such as comments, likes, and shares received on a post are good indicators of its popularity. However, each of these actions involves a different level of interaction and are assigned different weights in the Facebook newsfeed algorithm~\cite{kim2017like} with share receiving the highest weight and like the least. Hence, we do not aggregate these counts but analyze them separately. In order to account for the large difference in popularity of different news sites under consideration, we scale these counts of affinity metrics in the range of 0 to 1 and use the normalized values to determine the popularity of news posts.

\begin{figure}[!h]
    \centering
  \includegraphics[width=10cm, height=5cm]{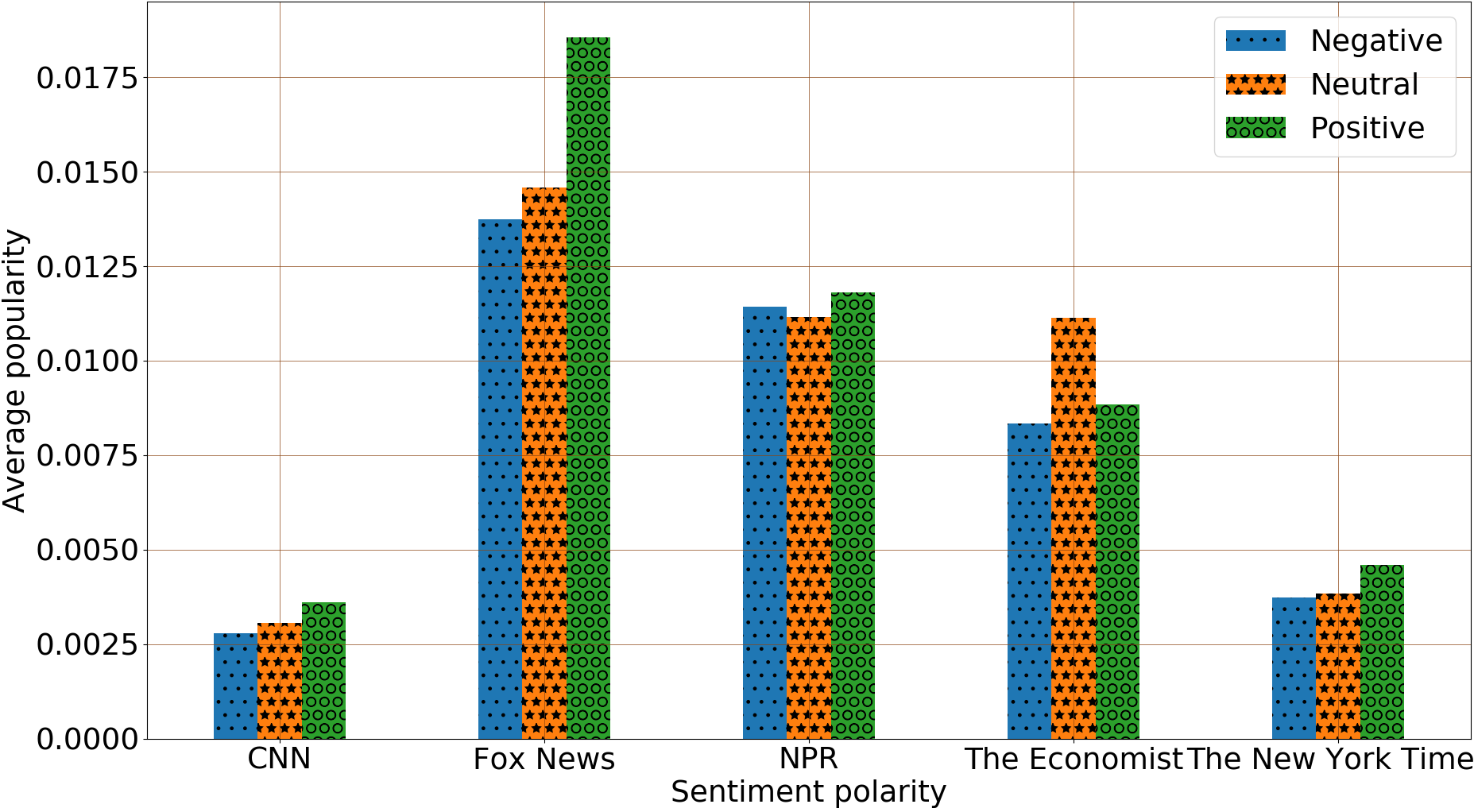}
    \caption{Likes}
    \label{fig:sa_likes}
\end{figure}

\begin{figure}[!h]
\minipage{0.5\textwidth}
  \includegraphics[width=7.5cm, height=5cm]{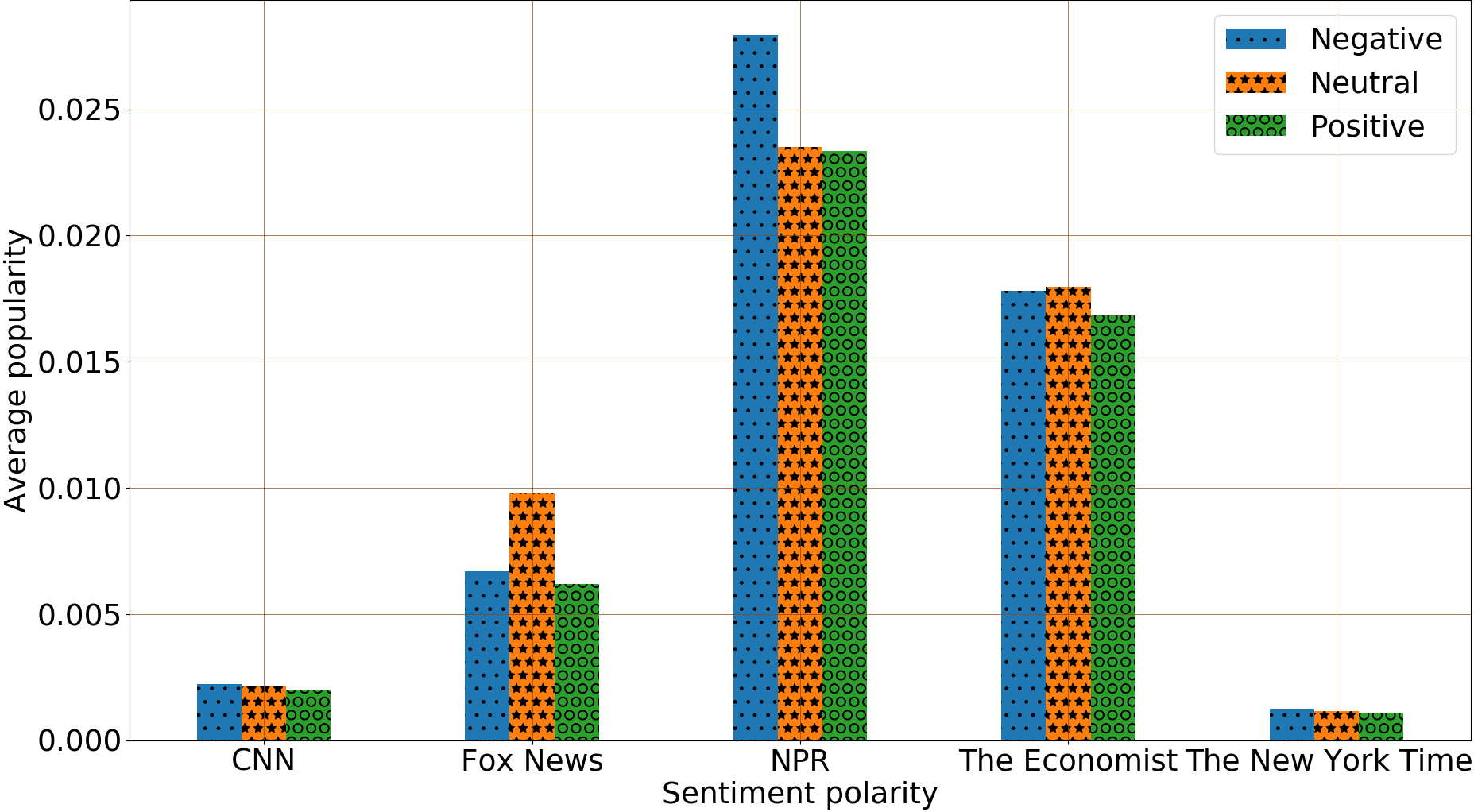}
  \caption{Comments}
  \label{fig:sa_comments}
\endminipage\hfill
\minipage{0.5\textwidth}%
  \includegraphics[width=7.5cm, height=5cm]{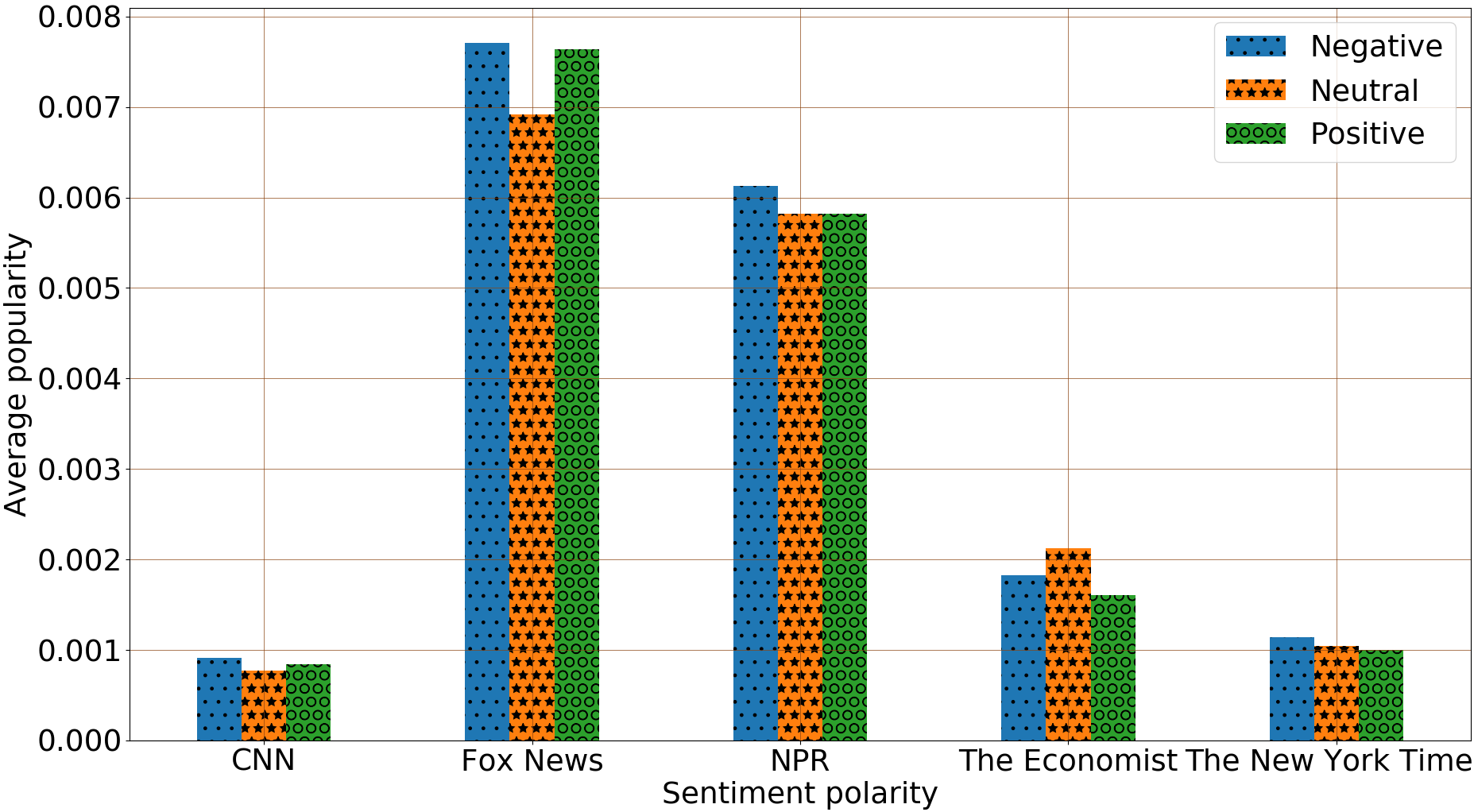}
  \caption{Shares}
  \label{fig:shares}
\endminipage
\end{figure}

We observe in Figure~\ref{fig:sa_likes}-\ref{fig:shares} that posts, which are either positive or negative, are more popular than the neutral ones in most of the cases. This suggests that news posts that are either positive or negative in tone tend to be more popular in social media. This finding is inline with observations made by Naveed et al.~\cite{naveed2011bad} stating that people are more attracted towards positive or negative news compared to neutral ones. Exceptions to this are Fox News only for comments and The Economist channel. One of the reasons for this is that The Economist reported a major fraction of the news related to money and lifestyle (refer to Figure~\ref{fig:all_cat}). The Economist also reported the highest number of money or business related news compared to other news channels. We have observed that these news were mostly reported facts and figures (usually neutral in sentiment), which leads to a higher number of reactions for neutral news.

Further, in relation to preference between positive and negative content, we observe that more likes are received for the positive posts whereas more comments and shares are received for the negative posts. It shows that the results agree with “Negativity bias” for actions that involve a greater level of engagement such as commenting and sharing but disagrees with when it comes to simpler actions such as liking the post. Trussler and Soroka~\cite{trussler2014negative} performed an eye tracking experiment to understand consumer demand for negative news frames and found a similar result. 
They found that participants “said”, they preferred
good news but in reality often chose negative news stories over positive ones. 
While it is apparent that negative news receives a greater level of engagement, it is important to understand whether users are engaging positively or negatively with content in order to design any plan to receive appropriate users’ opinions. We answer this question in the upcoming section.

\section{User Opinion Analysis}
\label{sec:anuo}
In addition to indicating the popularity of the post, user opinion (or comment) can provide a great deal of information about the tone of the audience, which can conclude whether the post is being perceived positively or negatively. 
To understand how users respond to posts of different sentiment polarities, we determine average sentiment polarity of comments received for each of the posts.
We show the relationship between average sentiment polarity of comments and  sentiment polarity of posts as follows:

\begin{figure}[H]
\minipage{0.5\textwidth}
  \includegraphics[width=7.5cm, height=5cm]{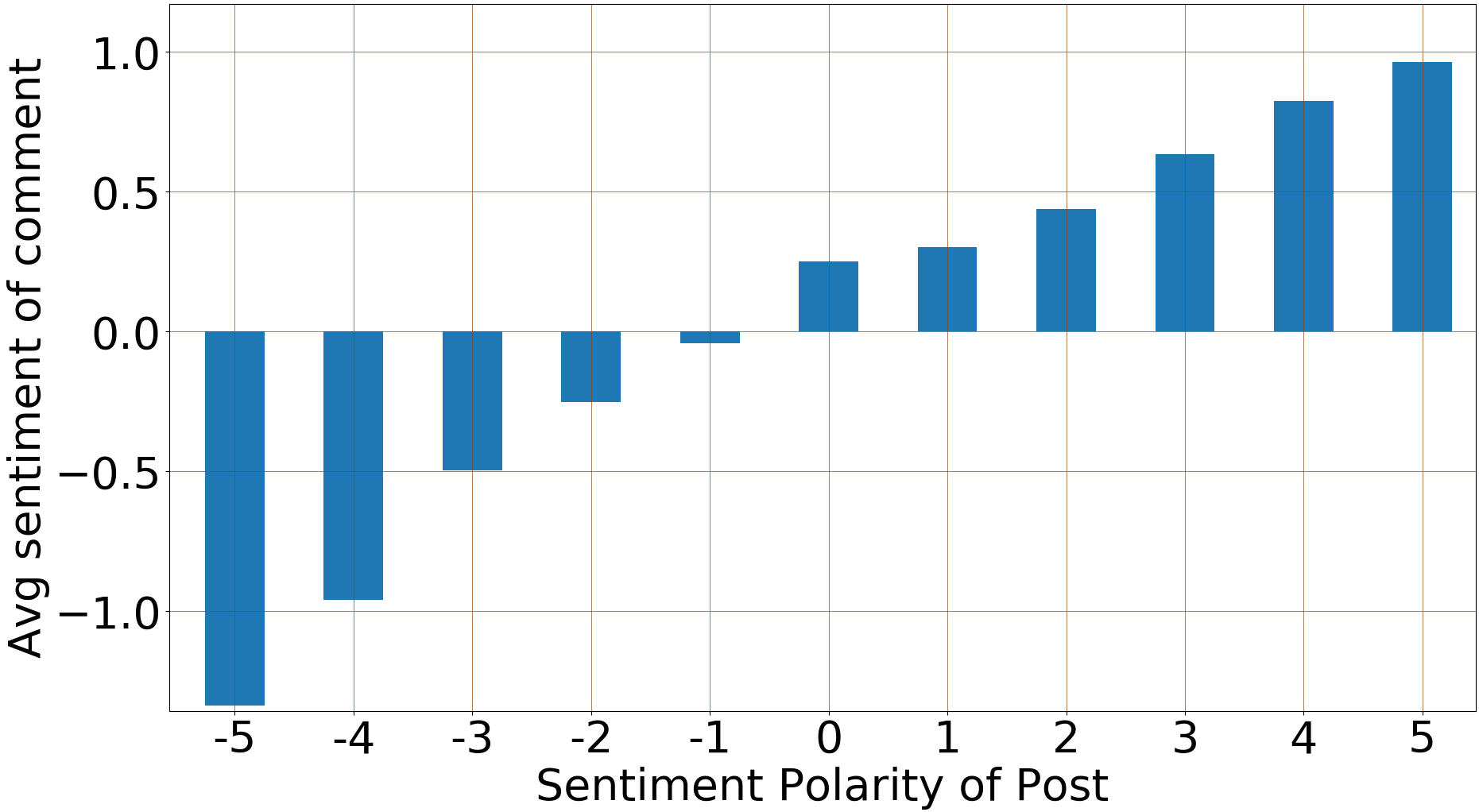}
  \caption{CNN}
  \label{fig:CNN_comm}
\endminipage\hfill
\minipage{0.5\textwidth}%
  \includegraphics[width=7.5cm, height=5cm]{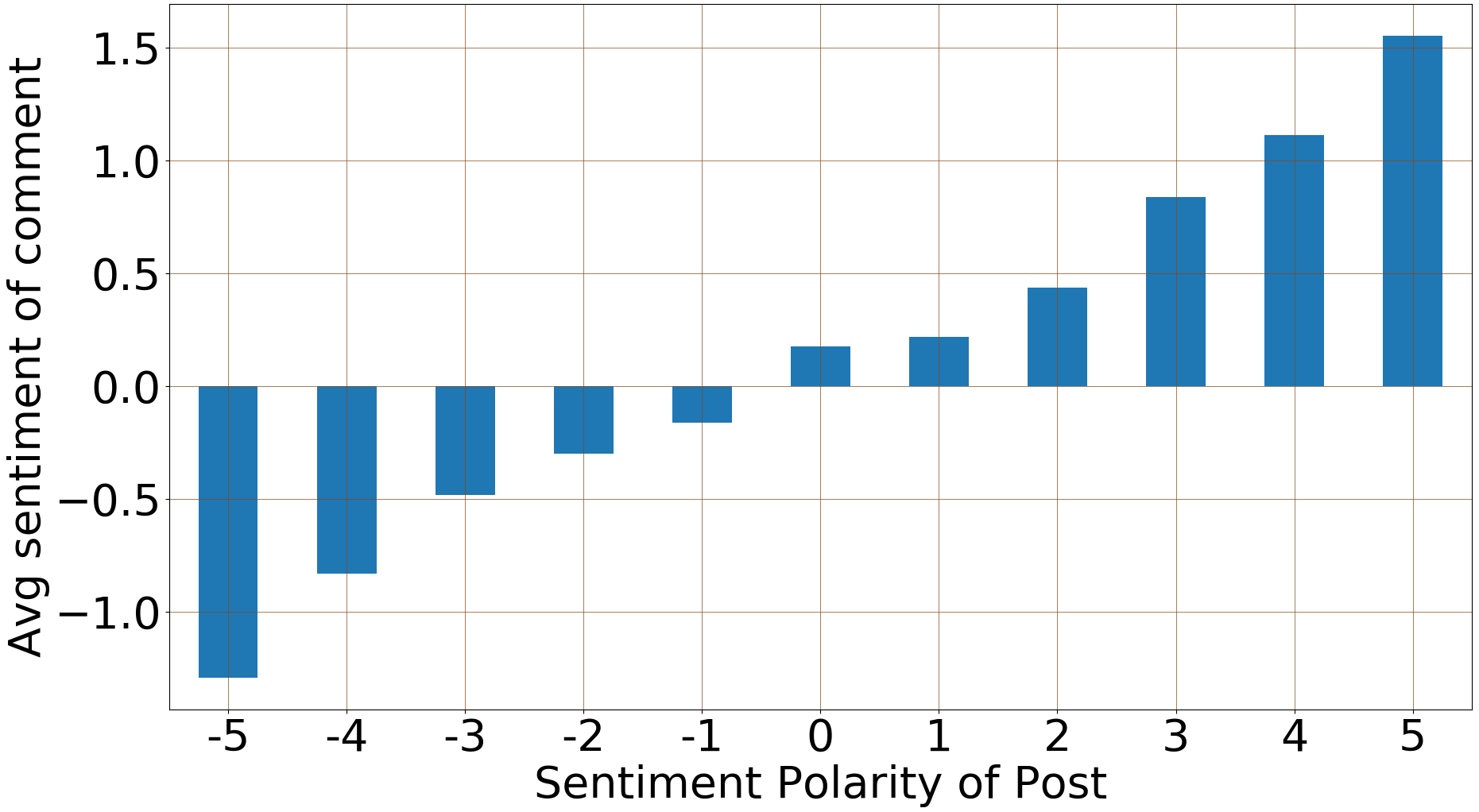}
  \caption{Fox News}
  \label{fig:fox_comm}
\endminipage
\end{figure}

\begin{figure}[!h]
\minipage{0.5\textwidth}
  \includegraphics[width=7.5cm, height=5cm]{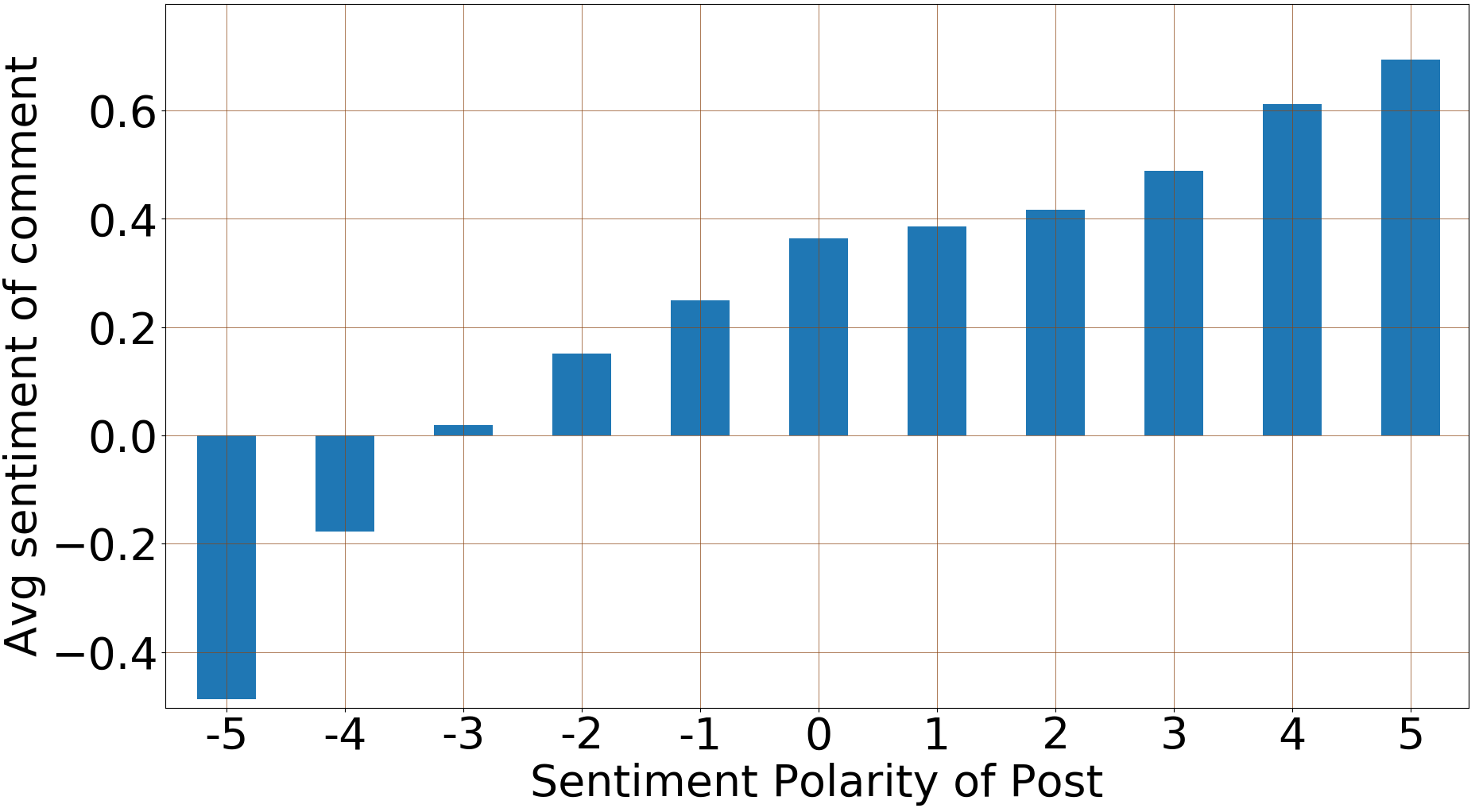}
  \caption{The Economist}
  \label{fig:econ_comm}
\endminipage\hfill
\minipage{0.5\textwidth}%
  \includegraphics[width=7.5cm, height=5cm]{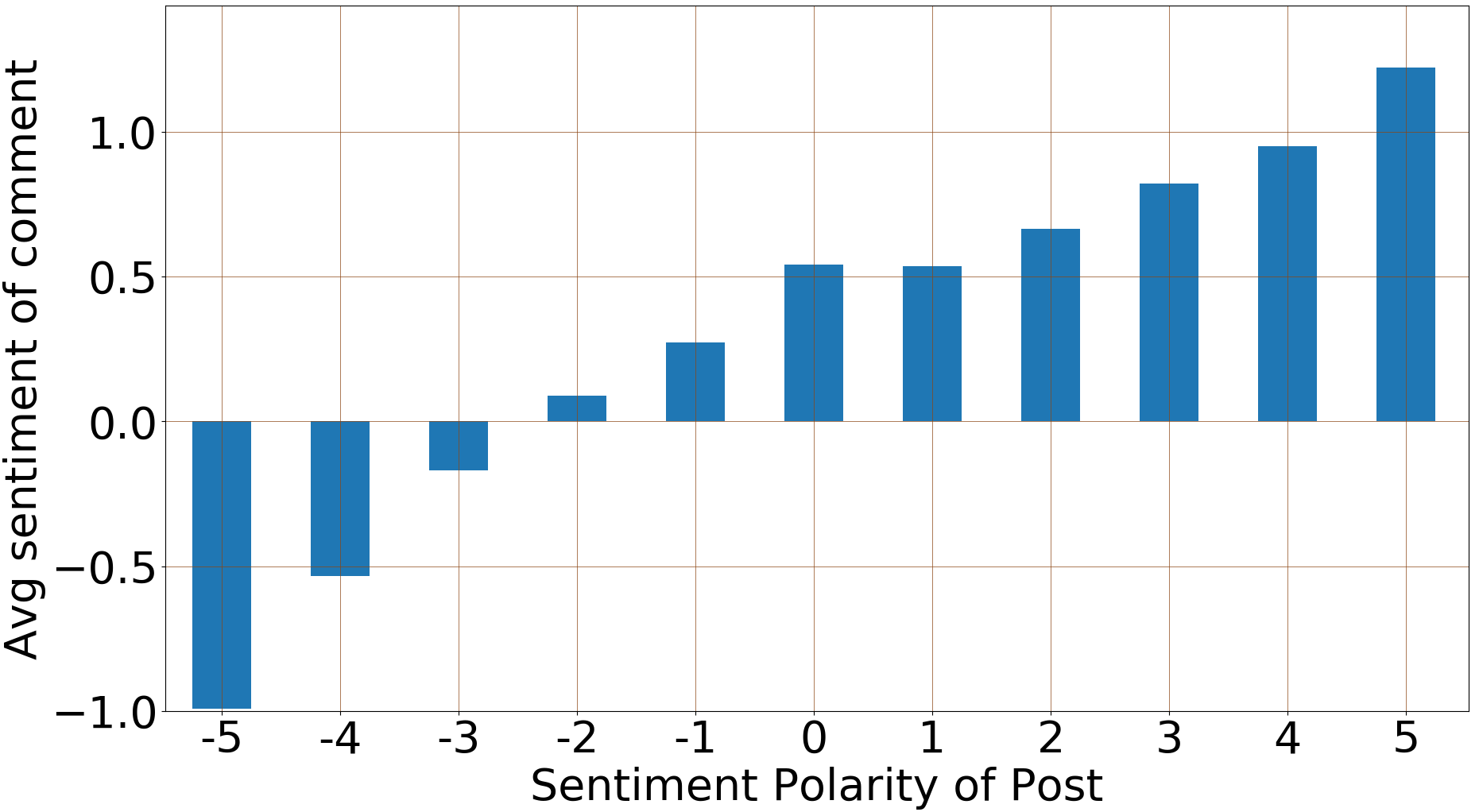}
  \caption{The New York Times}
  \label{fig:NYT_comm}
\endminipage
\end{figure}

\begin{figure}[!h]
\minipage{0.5\textwidth}
  \includegraphics[width=7.5cm, height=5cm]{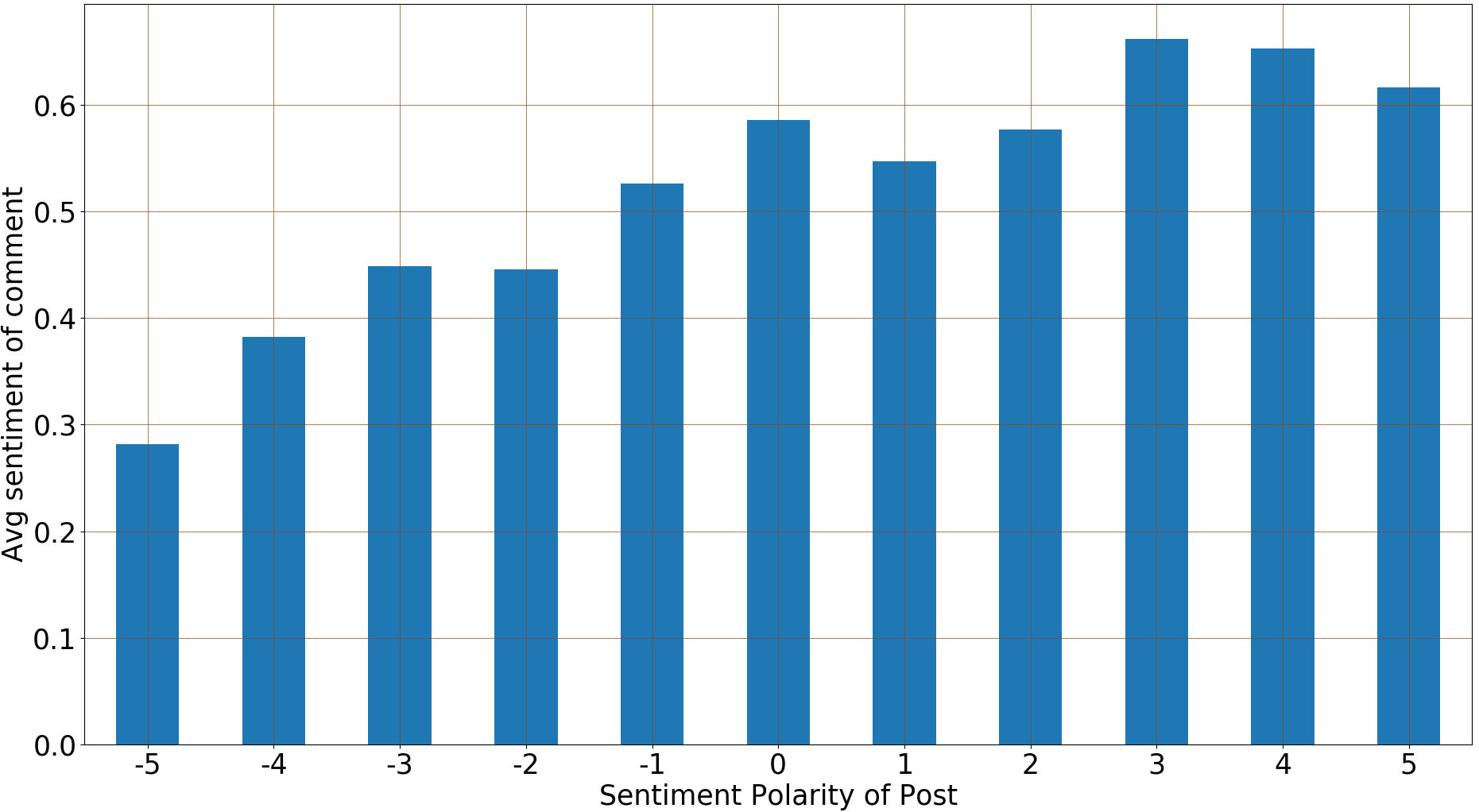}
    \caption{NPR}
    \label{fig:NPR_comm}
\endminipage\hfill
\minipage{0.5\textwidth}%

 \begin{center}
  \small
  \begin{tabular}{|l|l|}
\hline
\hline
 \thead{News Channel} & \thead{Correlation} \\
 \hline\hline
CNN &0.97 \\
 \hline
Fox News &0.98 \\
 \hline
The Economist &0.95 \\
 \hline
NYT  &0.97\\
 \hline
 NPR &0.93\\
 \hline
\end{tabular}
\captionsetup{type=table}
 \caption{Correlation between post sentiment and comment sentiment}
\label{table:sa_com_corr}
 \end{center}

\endminipage
 \end{figure}

We observe from all the five channels that as posts become more and more positive, comments also become increasingly positive. As can also be seen in Table~\ref{table:sa_com_corr} that there is a strong correlation between the sentiment polarity of comments and the sentiment polarity of posts. Comments of all the three types of channels have high sentiment correlation with the posts, and among all the channels TV based channels show the highest correlation. To validate our findings, we perform p-test~\cite{dahiru2008p}, which shows correlations between posts and comments sentiments are significant at p < .05. We can thus infer that the posts written with varying levels of sentiment polarity prompt different reactions from users. A high correlation between post sentiment and comment sentiment suggests that measures of sentiment polarity of posts can be used to correct for biases that occur when aggregating comments from various channels for tasks such as opinion mining, opinion summarization, real-world outcome prediction, etc.

However, the polarity for which a post starts attracting negative comments varies based on the medium of the channel. While Facebook pages of TV based channels, on an average, attract negative comments for negative posts and vice versa, comments for print media based channels do not become negative until posts become strongly negative in tone (i.e. sentiment score less than -3). It is interesting to note that the average comment sentiment polarity of NPR, which posts the highest proportion of positive content amongst all the channels, remains positive irrespective of the polarity of post. It must be recalled  from Section~\ref{sec:polnp} that posts by TV based news channels were predominantly negative whereas those by print media and radio based channels were predominantly positive with radio based channel having the highest percentage of positive posts. This suggests that sentiment expressed in the comments is not only strongly influenced by the polarity of that particular post but also by users' opinion about the channel posting the news. The user opinion about the news channel is in turn shaped by the whether the majority of the post messages have a positive or negative tilt. That is, Facebook pages of TV news channels which mostly post negative content attract more negative comments whereas channels that are positive in tone like print media and radio attract fewer negative comments.

\section{Temporal Analysis}
\label{sec:senti_tempana}
In this Section, we analyze the polarity of news posts temporally. We investigate how the polarity varies over the years. We also investigate whether posts of certain polarity drastically increase or decrease in particular months or days of the week. A common time frame from December 2014 to December 2016 is considered for the analysis. We present the results as follows:

\begin{figure}[!h]
\minipage{0.5\textwidth}
  \includegraphics[width=7.5cm, height=5.4cm]{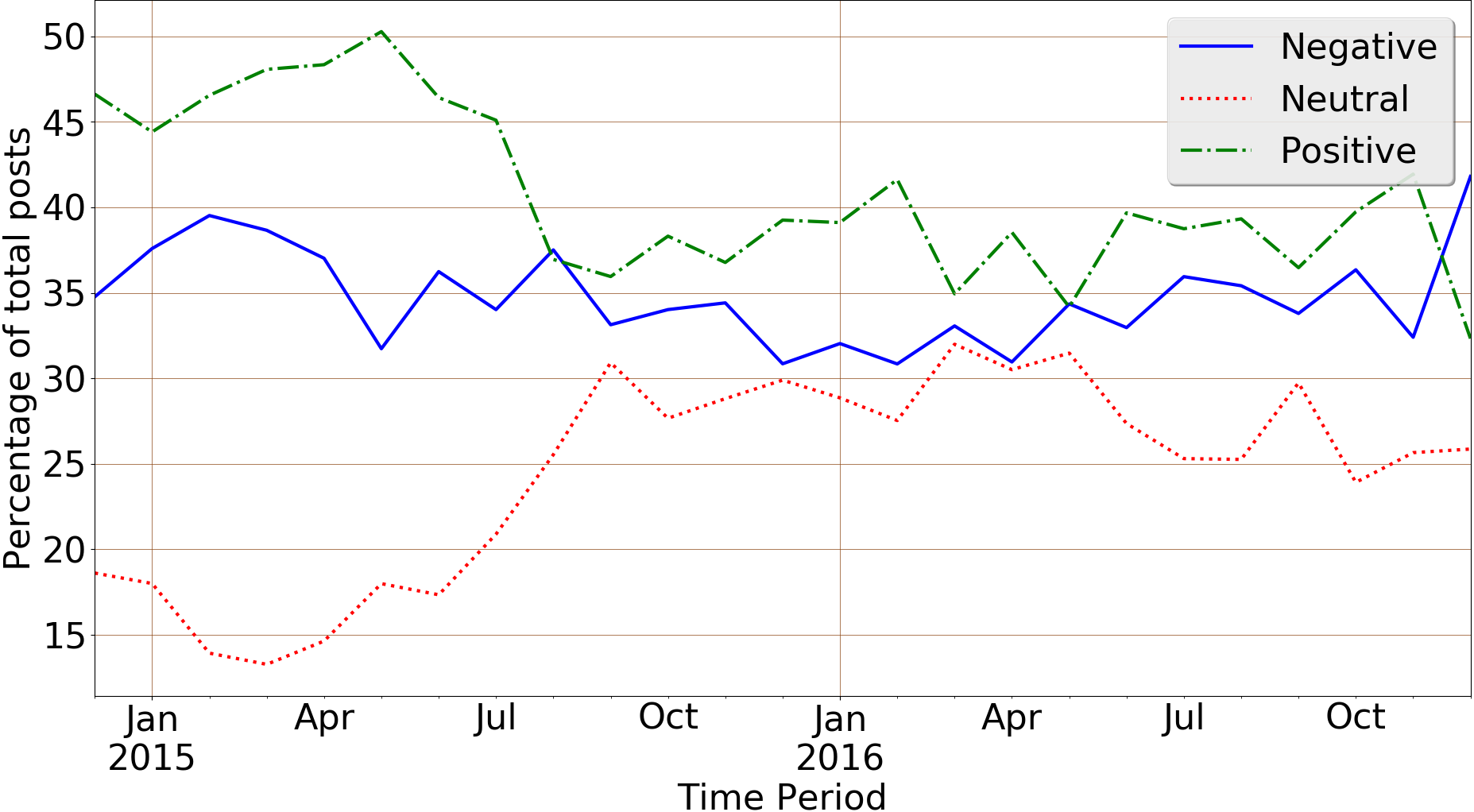}
  \caption{The Economist}
  \label{fig:econ_yearly}
\endminipage\hfill
\minipage{0.5\textwidth}%
 \includegraphics[width=7.5cm, height=5.4cm]{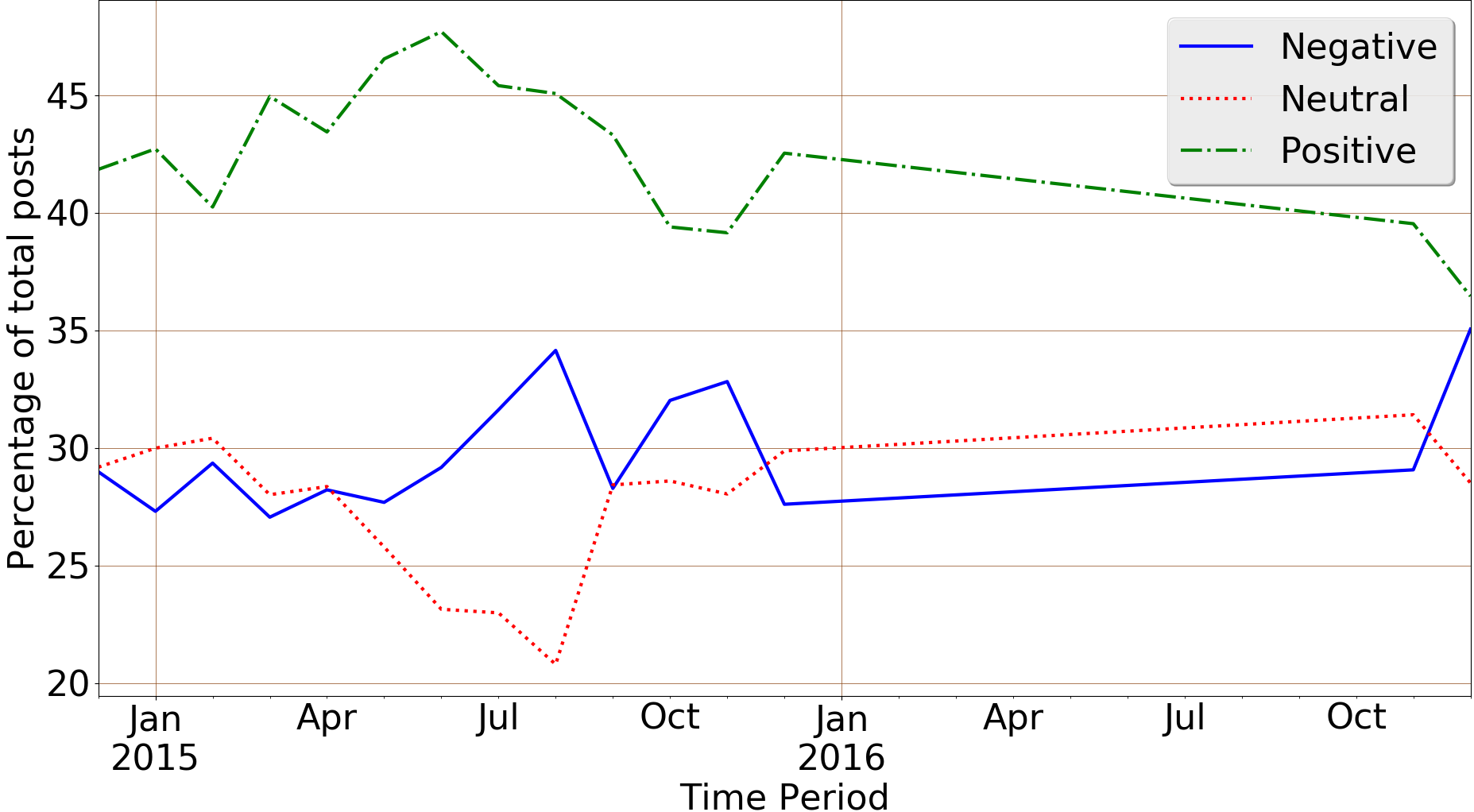}
  \caption{NPR}
  \label{fig:npr_yearly}
\endminipage
\end{figure}

\begin{figure}[!h]
\minipage{0.5\textwidth}
  \includegraphics[width=7.5cm, height=5.4cm]{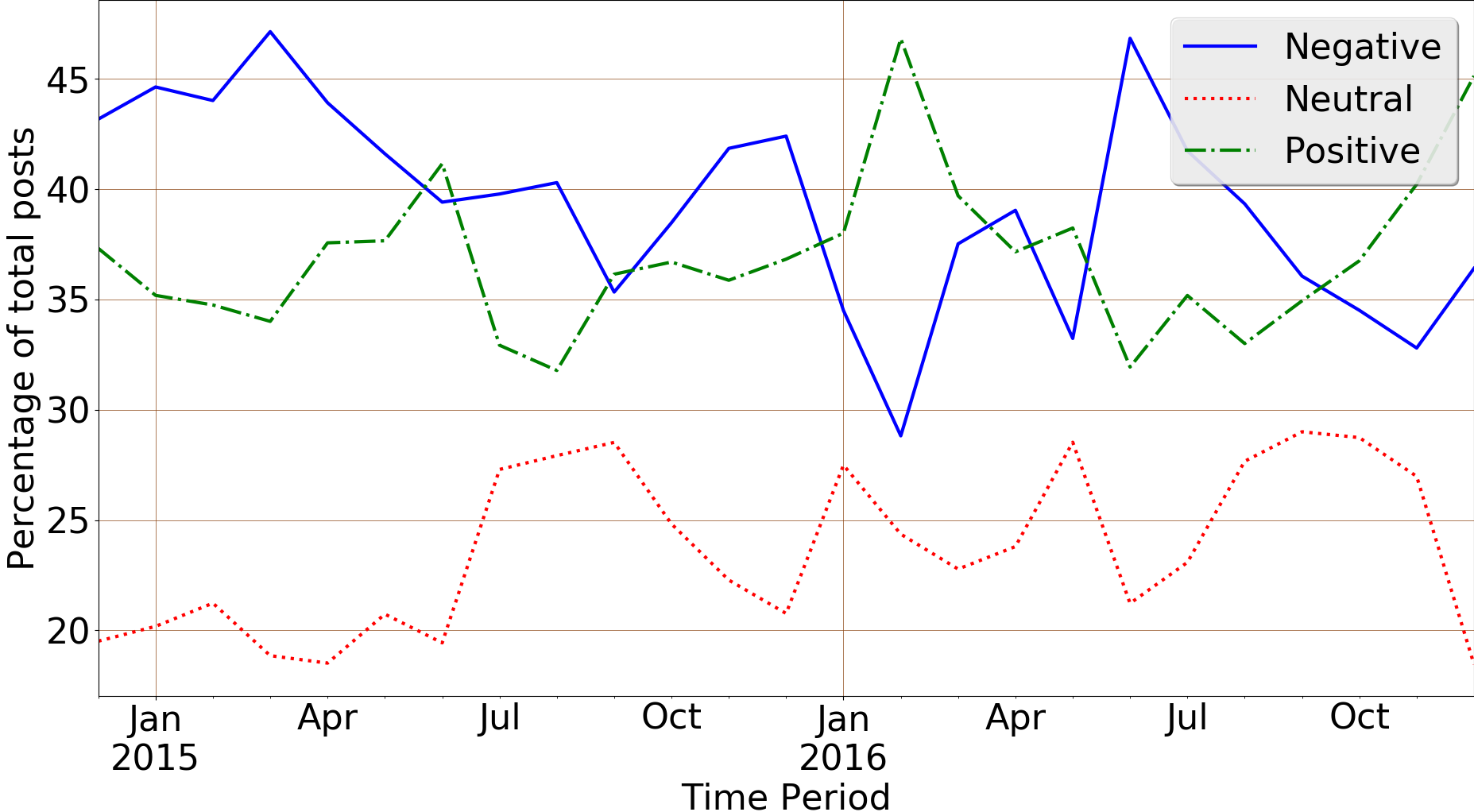}
  \caption{Fox News}
  \label{fig:fox_yearly}
\endminipage\hfill
\minipage{0.5\textwidth}%
  \includegraphics[width=7.5cm, height=5.4cm]{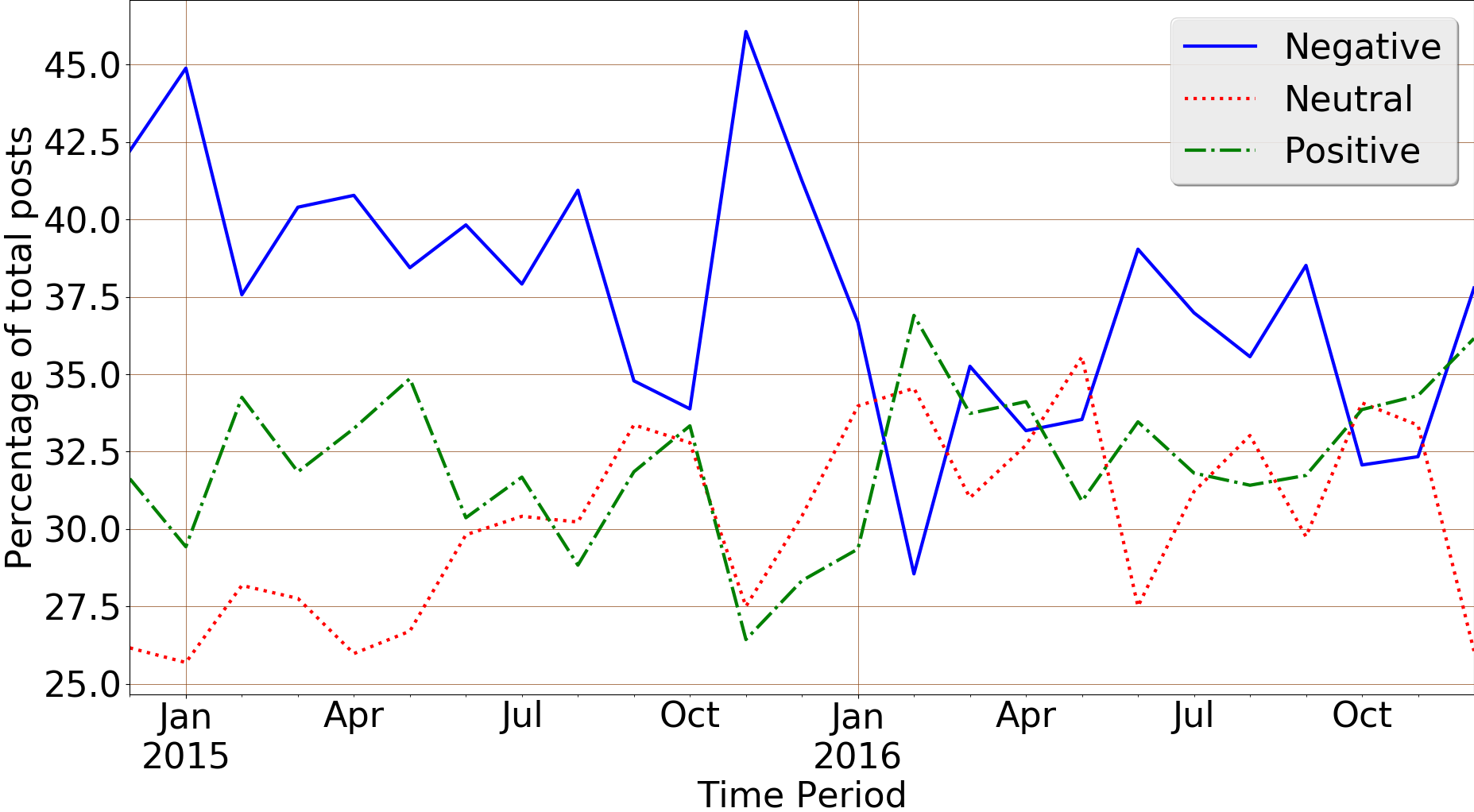}
  \caption{CNN}
  \label{fig:cnn_yearly}
\endminipage
\end{figure}

Figures~\ref{fig:fox_yearly}-\ref{fig:npr_yearly} show the polarity of the news articles over the period of years. We observe that the behavior of television, print and radio based channels remain same as that revealed in Section~\ref{sec:polnp}, i.e. negative sentiment dominates in the television based channels, positive sentiment dominates in the radio and print media based ones majority of the time. 

We observe that, over the time, Facebook pages of print based media channels (Fig.~\ref{fig:econ_yearly}) show a gradual decrease in the percentage of positive posts  while neutral posts increase and  negative ones remain almost constant. As can also be seen in Figure~\ref{fig:econ_yearly} and \ref{fig:npr_yearly}, there is a peak of positive sentiment during the months of April to July 2015. One of the reasons for this is that two big headlines about Obamacare and same-sex marriage were in the trending news during that time. A few example news of these headlines are as follows: (1). The Supreme Court has ruled on Obamacare subsidies; (2). US supreme court declares same-sex marriage legal. These big news headlines lead to a sentiment peak in Figure~\ref{fig:econ_yearly} and \ref{fig:npr_yearly}, which is also in line with our analysis in Section~\ref{sec:headlinevsniche} that report that news channels slightly generate more positive news for big headlines compared to negative news.

Both the Television based channels, CNN and FOX exhibit a slight increase in the positive news over the negative news in the year 2016. Such instances of graphs of same-medium news sites (Fig.~\ref{fig:fox_yearly} and \ref{fig:cnn_yearly}) showing similar trends suggest that they behave similarly and respond similarly to external events. 
One of the primary reasons for slightly dip in negative sentiment and a rise in positive sentiment is a big headline, namely US presidential election. These TV based channels are somewhat biased towards a party and generate positive news about that party. Another big headline news was Nobel peace prize award to Colombian president. These headlines are one of the reasons for slightly dip in negative news and rise in positive sentiment.

We also analyze the news posts over the months and weeks but we do not notice considerable variations in their polarities. On inspecting sentiment distribution over the months, we do not observe any consistent and significant change for certain months of the year. Even the weekly analysis does not reveal any significant change in the distribution except for a slight increase in positive posts on weekends. One of the reasons for this is that news channels have posted slightly more entertainment, lifestyle, and sports news during the weekend. This posting behavior of news channels leads to a slight increase in positive posts on weekends. Moreover, it is noted that the dominant sentiment in both the monthly and weekly analysis is also similar to the behavior of the channels observed in previous Section~\ref{sec:polnp}. Thus, by analyzing the sentiment temporally, we can conclude that, on average, the specific characteristics observed in Section~\ref{sec:polnp} are exhibited consistently across the weeks, months and years.

\section{Conclusion}
\label{sec:sa_cnfw}
In this paper, we conducted an extensive analysis on social media news channels from three types of news information sources to analyze the sentiment of the news generated by these channels and its effect on users' reactions. We characterized the news in different categories to uncover the distribution of the news posts and their sentiment polarity across categories. 
Our analysis revealed that sentiment of the news posted by different types of social media channels is dependent on the medium through which these channels traditionally disseminated the news. We also investigated popularity of news with different polarity to get insight into the type of news that attract lots of users' attention. Interestingly, we found that news with positive or negative sentiment receive lots of users’ attention. We also found that users' opinion depend on the sentiment of news articles and the type of information sources. Finally, we performed temporal analysis to understand how news posting behaviour of social media channels evolve over a period of time. Future work can look at actual online news articles from different types of online news channels to investigate the differences in the news articles and to compare these articles with social media news posts.



\bibliographystyle{elsarticle-num}

\bibliography{sample}

\begin{thebibliography}{10}
\bibitem{d2013dont}
K.~D’Costa, Dont read the comments!(why do we read the online comments when
  we know theyll be bad?), Scientific American.
\bibitem{moosa2014comment}
T.~Moosa, Comment sections are poison: handle with care or remove them, The
  Guardian.
\end{thebibliography}

\end{document}